\documentclass[12pt,english]{article}
\usepackage{times}
\usepackage[T1]{fontenc}
\usepackage{geometry}
\usepackage{graphicx}
\usepackage{setspace}
\usepackage{amssymb}

\makeatletter

\providecommand{\tabularnewline}{\\}

\usepackage{bm}

\usepackage{babel}
\makeatother
\begin{document}

\section*{\noindent Co-Design of Low-Profile Linear Microstrip Arrays with
Wide-Band Spatial Filtering Capabilities}

\noindent {\scriptsize ~}{\scriptsize \par}

\noindent \vfill

\noindent A. Benoni,$^{\left(1\right)}{}^{\left(2\right)}$ M. Salucci,$^{\left(1\right)}{}^{\left(2\right)}$
\emph{Senior Member}, \emph{IEEE}, and A. Massa,$^{\left(1\right)}{}^{\left(2\right)}{}^{\left(3\right)}{}^{\left(4\right)}{}^{\left(5\right)}$
\emph{Fellow, IEEE}

\noindent \vfill

\noindent {\scriptsize ~}{\scriptsize \par}

\noindent {\small $^{(1)}$} \emph{\footnotesize ELEDIA Research Center}
{\footnotesize (}\emph{\footnotesize ELEDIA}{\footnotesize @}\emph{\footnotesize UniTN}
{\footnotesize - University of Trento)}{\footnotesize \par}

\noindent {\footnotesize DICAM - Department of Civil, Environmental,
and Mechanical Engineering}{\footnotesize \par}

\noindent {\footnotesize Via Mesiano 77, 38123 Trento - Italy}{\footnotesize \par}

\noindent \textit{\emph{\footnotesize E-mail:}} {\footnotesize \{}\emph{\footnotesize arianna.benoni,
marco.salucci, andrea.massa}{\footnotesize \}@}\emph{\footnotesize unitn.it}{\footnotesize \par}

\noindent {\footnotesize Website:} \emph{\footnotesize www.eledia.org/eledia-unitn}{\footnotesize \par}

\noindent {\tiny ~}{\tiny \par}

\noindent {\footnotesize $^{(2)}$} \emph{\footnotesize CNIT - \char`\"{}University
of Trento\char`\"{} ELEDIA Research Unit }{\footnotesize \par}

\noindent {\footnotesize Via Mesiano 77, 38123 Trento - Italy }{\footnotesize \par}

\noindent {\footnotesize E-mail: \{}\emph{\footnotesize arianna.benoni,
marco.salucci, andrea.massa}{\footnotesize \}}\emph{\footnotesize @unitn.it}{\footnotesize \par}

\noindent {\footnotesize Website:} \emph{\footnotesize www.eledia.org/eledia-unitn}{\footnotesize \par}

\noindent {\tiny ~}{\tiny \par}

\noindent {\footnotesize $^{(3)}$} \emph{\footnotesize ELEDIA Research
Center} {\footnotesize (}\emph{\footnotesize ELEDIA}{\footnotesize @}\emph{\footnotesize UESTC}
{\footnotesize - UESTC)}{\footnotesize \par}

\noindent {\footnotesize School of Electronic Science and Engineering,
Chengdu 611731 - China}{\footnotesize \par}

\noindent \textit{\emph{\footnotesize E-mail:}} \emph{\footnotesize andrea.massa@uestc.edu.cn}{\footnotesize \par}

\noindent {\footnotesize Website:} \emph{\footnotesize www.eledia.org/eledia}{\footnotesize -}\emph{\footnotesize uestc}{\footnotesize \par}

\noindent {\tiny ~}{\tiny \par}

\noindent {\footnotesize $^{(4)}$} \emph{\footnotesize ELEDIA Research
Center} {\footnotesize (}\emph{\footnotesize ELEDIA@TSINGHUA} {\footnotesize -
Tsinghua University)}{\footnotesize \par}

\noindent {\footnotesize 30 Shuangqing Rd, 100084 Haidian, Beijing
- China}{\footnotesize \par}

\noindent {\footnotesize E-mail:} \emph{\footnotesize andrea.massa@tsinghua.edu.cn}{\footnotesize \par}

\noindent {\footnotesize Website:} \emph{\footnotesize www.eledia.org/eledia-tsinghua}{\footnotesize \par}

\noindent {\tiny ~}{\tiny \par}

\noindent {\small $^{(5)}$} {\footnotesize School of Electrical Engineering}{\footnotesize \par}

\noindent {\footnotesize Tel Aviv University, Tel Aviv 69978 - Israel}{\footnotesize \par}

\noindent \textit{\emph{\footnotesize E-mail:}} \emph{\footnotesize andrea.massa@eng.tau.ac.il}{\footnotesize \par}

\noindent {\footnotesize Website:} \emph{\footnotesize https://engineering.tau.ac.il/}{\footnotesize \par}

\noindent \vfill

\noindent \textbf{\emph{This work has been submitted to the IEEE for
possible publication. Copyright may be transferred without notice,
after which this version may no longer be accessible.}}

\noindent \vfill

\newpage
\section*{Co-Design of Low-Profile Linear Microstrip Arrays with Wide-Band
Spatial Filtering Capabilities}

\noindent ~

\noindent ~

\noindent ~

\begin{flushleft}A. Benoni, M. Salucci, and A. Massa\end{flushleft}

\vfill

\begin{abstract}
\noindent The design of low-profile linear microstrip arrays with
wide-band spatial filtering capabilities is dealt with. An innovative
architecture, leveraging the angular selectivity of offset stacked
patch (\emph{OSP}) radiators, is proposed to implement phased arrays
(\emph{PA}s) with inter-element spacing larger than half-wavelength
that feature remarkable grating lobes (\emph{GL}s) suppression properties
and an enhanced gain within a non-negligible down-looking scanning
angular range. The \emph{PA} layout is then obtained by optimizing
the optimal \emph{micro-scale} geometrical descriptors of the radiating
elements so that the \emph{macro-scale} electromagnetic (\emph{EM})
features of the arising finite-size \emph{PA} fulfill the user-defined
requirements. A set of numerical test cases, concerned with a variation
of the array size and its polarization, is presented to assess the
capabilities, the flexibility, and the potentialities of the proposed
spatial filtering technique (\emph{SFT}) also in comparison with competitive
state-of-the-art alternatives. The performance of a printed circuit
board (\emph{PCB})-manufactured prototype are experimentally assessed,
as well.

\noindent \vfill
\end{abstract}
\noindent \textbf{Key words}: Phased Array (\emph{PA}) Design; Spatial
Filtering Techniques (\emph{SFT}s); Grating Lobe (\emph{GL}) Suppression;
Offset Stacked Patches (\emph{OSP}s); System-by-Design (\emph{SbD}).

\newpage
\section{Introduction and Motivation}

In modern communication and radar systems, phased arrays (\emph{PA}s)
have often to operate in complex environments where multiple radiating
infrastructures, wireless devices, and sensors coexist \cite{Yepes 2020.b}-\cite{Ramaccia 2022}.
Therefore, it is paramount to avoid interferences between the different
systems. Towards this end, many practical scenarios require that the
gain pattern of \emph{PA}s fulfills a user-defined mask only within
a limited scanning region depending on the targeted application \cite{Yepes 2020.b}\cite{Kim 2015}.
For instance, interference mitigation/immunity of base transceiver
stations (\emph{BTS}s) is often improved by employing down-looking
radiation patterns pointing towards the mobile end-users to yield
an enhanced capacity and received signal strength \cite{Yepes 2020.b}.
While a mechanical tilting of the \emph{BTS} antenna could in principle
provide the desired performance, an electronic steering of the main
beam direction is currently a preferred choice since it enables an
easier reconfigurability to the changing operative/environmental conditions
\cite{Yepes 2020.b}-\cite{Mailloux 2018}. On the other hand, the
need for high-directivity and low-cost antennas, fulfilling more and
more challenging constraints when moving from current to next-generation
standards \cite{Cox 2021}\cite{Hataria 2021}, often implies the
adoption of \emph{PA}s with inter-element spacing larger than the
conventional half-wavelength one without recurring to a more complex
and expensive back-end circuitry and beam-forming network. Moreover,
such a solution allows one to reduce the mutual coupling (\emph{MC})
among the elementary radiators \cite{Ramaccia 2022}\cite{Balanis 2016}-\cite{Mailloux 2018},
as well. However, the {}``price to pay'' is the unavoidable insurgence
of grating lobes (\emph{GL}s) in the array far-field (\emph{FF}) pattern
that correspond to spurious radiations towards above-the-horizon directions,
while scanning the beam in elevation within the down-looking field-of-view
(\emph{FoV}) \cite{Ramaccia 2022}. Such undesired phenomena cause
a loss of both power and directivity \cite{Blanco 2014}\cite{Blanco 2016}
and they may generate unacceptable interferences towards geostationary
satellites \cite{Ramaccia 2022}. Therefore, the implementation of
spatial filtering techniques (\emph{SFT}s) to suppress undesired radiation
towards unintended angular directions is of great interest \cite{Yepes 2020.b}-\cite{Ramaccia 2022}\cite{Blanco 2014}-\cite{Iqbal 2019}.
Effective solutions have been recently proposed where the elementary
radiators have been designed to radiate asymmetric patterns having
higher gains within the angular scanning region of the array and lower
gains where the \emph{GL}s occur \cite{Yepes 2020.b}\cite{Kim 2015}\cite{Yepes 2020.a}.
Thanks to such an {}``angular selectivity'' of the radiators, it
is thus possible to spatially filter out the \emph{GL}s of the \emph{}array
pattern\emph{,} while enhancing the array gain only within the desired
\emph{FoV}. For instance, asymmetric active element patterns have
been yielded in \cite{Yepes 2020.b}\cite{Yepes 2020.a} by means
of skewed dipoles (\emph{SD}s) loaded with parasitic strips. Differently,
electromagnetic band-gap (\emph{EBG}) structures have been successfully
exploited in \cite{Kim 2015} to tilt the radiation patterns of standard
microstrip dipoles by properly tailoring the surface current distribution
induced on their ground planes. Other promising solutions are those
based on leaky-wave antennas (\emph{LWA}s) built by placing a partially
reflective surface (\emph{PRS}) at a proper distance from the antenna
aperture \cite{Blanco 2014}\cite{Blanco 2016}. Alternatively, metasurface
(\emph{MTS})-based lenses/domes have been recently designed to manipulate
the \emph{FF} pattern of the underlying \emph{PA}s with anomalous
refraction phenomena for mitigating the insurgence of \emph{GL}s caused
by an inter-element spacing larger than a half wavelength \cite{Ramaccia 2022}\cite{Jiang 2017}. 

\noindent Although effective, the \emph{SFT} implementations in the
state-of-the-art literature are generally not low-profile. For instance,
with reference to the illustrative sketch in Fig. 1, the antenna thickness
turns out to be $T_{SD}=0.7$ {[}$\lambda_{c}${]} in \cite{Yepes 2020.b},
$T_{EBG}=0.84$ {[}$\lambda_{c}${]} in \cite{Kim 2015}, $T_{PRS}=0.46$
{[}$\lambda_{c}${]} in \cite{Blanco 2016}, and $T_{MTS}=1.0$ {[}$\lambda_{c}${]}
in \cite{Ramaccia 2022}, $\lambda_{c}$ being the free-space wavelength
at the central working frequency $f_{c}$. Moreover, some solutions
are competitive only in narrow bandwidths (e.g., the fractional bandwidth
of the \emph{PRS}-based \emph{SFT} in \cite{Blanco 2016} is $FBW=1.74\%$),
and/or support limited \emph{FoV}s (e.g., the maximum scan angle in
\cite{Blanco 2014} is $8.6$ {[}deg{]}), or they are designed for
a fixed down-looking direction \cite{Kim 2015}.

\noindent This work is aimed at presenting a simple, but rather effective,
implementation of low-profile (i.e., $T_{OSP}<0.15$ {[}$\lambda_{c}${]}
- Fig. 1) linear microstrip arrays with wide-band spatial filtering
capabilities. The proposed \emph{PA} architecture exploits - for the
first time to the best of the authors' knowledge - offset stacked
patches (\emph{OSP}s) \cite{Rajo-Iglesias 2002}-\cite{Rajo-Iglesias 2003},
to design low-profile \emph{PA}s with a remarkable \emph{GL}s suppression
and enhanced gain within a non-negligible down-looking angular scanning
range. In order to synthesize robust and reliable array layouts, a
\emph{co-design} strategy \cite{Oliveri 2017b} is introduced to take
into account the effects of \emph{MC} in the synthesis process for
faithfully predicting the \emph{FF} features of the arising finite-size
\emph{PA}. 

\noindent The main novelties of this work over the existing literature
consist, to the best of the authors' knowledge, in (\emph{i}) the
exploitation of \emph{OSP} radiators, which are traditionally employed
for bandwidth/gain enhancement \cite{Rajo-Iglesias 2002}-\cite{Katyal 2017.a}
and/or \emph{MC} reduction \cite{Katyal 2017.b}, as building blocks
of low-profile microstrip \emph{PA}s with spatial filtering capabilities
and (\emph{ii}) the solution of the \emph{PA} synthesis problem at
hand with an innovative \emph{co-design} strategy where the \emph{micro-scale}
descriptors of the elementary radiators are derived by directly optimizing
the \emph{macro-scale} electromagnetic (\emph{EM}) features of the
corresponding \emph{PA}. As for this latter item (\emph{ii}), the
optimization problem is formulated within the System-by-Design (\emph{SbD})
framework \cite{Oliveri 2017}-\cite{Salucci 2021} to effectively
cope with the computational burden caused by the need for repeated
full-wave (\emph{FW}) simulations of the whole finite size array instead
of the single stand-alone \emph{OSP} radiator. 

\noindent The outline of the paper is as follows. Section \ref{sec:Design-Concept-and-Principle}
illustrates the proposed design concept and the working principle
of the \emph{OSP}-based architecture, while Section \ref{sec:Synthesis-Methodology}
describes the proposed co-design \emph{}strategy within the \emph{SbD}
optimization framework. A representative set of numerical examples,
concerned with the design of \emph{PA}s of different sizes and polarizations,
is discussed in Sect. \ref{sec:Numerical-Assessment} to illustrate
the design/method features as well as to assess the effectiveness,
the flexibility, and the high modularity of the proposed \emph{SFT}
solution also in comparison with competitive state-of-the-art alternatives.
The \emph{OSP}-based design concept is also experimentally verified
with a \emph{PCB}-manufactured prototype operating in the $\left[24,\,28\right]$
{[}GHz{]} \emph{mm}-wave band (Sect. \ref{sec:Experimental-Validation}).
Eventually (Sect. \ref{sec:Conclusions}), some conclusions and final
remarks are drawn.

\section{\noindent Design Concept and Working Principle \label{sec:Design-Concept-and-Principle}}

Let us consider a linear array of $N$ microstrip patches centered
at \{$z_{n}=\left[\gamma-\left\lfloor \frac{N}{2}\right\rfloor +\left(n-1\right)\right]\times d$;
$n=1,...,N$\}, where $d$ is the uniform inter-element distance and
$\gamma=\frac{1}{2}$ if $\mathrm{mod}\left(N,\,2\right)=0$, $\gamma=0$
otherwise, $\mathrm{mod}\left(\,.\,\right)$ and $\left\lfloor \,.\,\right\rfloor $
being the modulo and the floor operators, respectively {[}Fig. 2(\emph{a}){]}.
The \emph{FF} electric field pattern radiated by the array is\begin{equation}
\mathbf{E}\left(\theta,\,\varphi\right)=\sum_{q=\left\{ \theta,\,\varphi\right\} }E_{q}\left(\theta,\,\varphi\right)\widehat{\mathbf{q}}\label{FF Pattern}\end{equation}
where\begin{equation}
E_{q}\left(\theta,\,\varphi\right)=\sum_{n=1}^{N}w_{n}\mathcal{E}_{nq}\left(\theta,\,\varphi\right)\exp\left[j\frac{2\pi}{\lambda}z_{n}\cos\left(\theta\right)\right]\label{eq:q-component}\end{equation}
is the $q$-th ($q=\left\{ \theta,\,\varphi\right\} $) \emph{FF}
pattern component, which is a function of the $n$-th ($n=1,...,N$)
complex excitation coefficient {[}$w_{n}\triangleq\alpha_{n}\exp\left(j\beta_{n}\right)${]}
and the $q$-th ($q=\left\{ \theta,\,\varphi\right\} $) active element
pattern of the $n$-th ($n=1,...,N$) radiator, $\mathcal{E}_{nq}\left(\theta,\,\varphi\right)$
\cite{Pozar 1994}, $\lambda$ being the free-space wavelength at
the working frequency. It is useful to notice that for an array of
isotropic elements {[}i.e., $\mathcal{E}_{nq}\left(\theta,\,\varphi\right)=\mathcal{E}_{n}\left(\theta,\,\varphi\right)=1$;
$n=1,...,N$; $q=\left\{ \theta,\,\varphi\right\} ${]}, the \emph{FF}
electric field pattern coincides with the array factor $AF\left(\theta,\,\varphi\right)$
($AF\left(\theta,\,\varphi\right)\triangleq\sum_{n=1}^{N}w_{n}\exp\left[j\frac{2\pi}{\lambda}z_{n}\cos\left(\theta\right)\right]$). 

\noindent Let also the array be steered along the pointing direction
$\theta_{s}$ with isophoric excitations, thus the amplitude and the
phase of the $n$-th ($n=1,...,N$) excitation turn out to be\begin{equation}
\begin{array}{l}
\alpha_{n}=1\\
\beta_{n}=-\frac{2\pi}{\lambda}z_{n}\cos\left(\theta_{s}\right).\end{array}\label{eq:excitations}\end{equation}
Subject to the condition that $\frac{\lambda}{2}<d\le\lambda$, a
grating lobe (\emph{GL}) appears in the (visible range of the) \emph{AF}
at the angular direction \cite{Balanis 2016}\begin{equation}
\theta_{GL}=\left\{ \begin{array}{ccc}
\arccos\left[\cos\left(\theta_{s}\right)+\frac{\lambda}{d}\right] & \textnormal{\,\,\, if\,\,\,} & 90<\theta_{s}<180\,\textnormal{[deg]}\\
\arccos\left[\cos\left(\theta_{s}\right)-\frac{\lambda}{d}\right] & \textnormal{\,\,\, if\,\,\,} & 0<\theta_{s}<90\,\textnormal{[deg]},\end{array}\right.\label{eq: GL}\end{equation}
which corresponds to a high side-lobe (\emph{SL}) in the \emph{FF}
electric field pattern radiated by the array (\ref{FF Pattern}),
that causes a loss of power and directivity. For instance, this is
the case of a \emph{BTS} array with the main beam pointed along a
fixed down-looking angular direction $90<\theta_{s}<180$ {[}deg{]}
that generates a significant \emph{SL} directed towards the sky {[}i.e.,
$\theta_{GL}<90$ {[}deg{]} - Fig. 2(\emph{a}){]}, thus a potential
interference with a satellite \cite{Yepes 2020.b}-\cite{Ramaccia 2022}\cite{Ye 2019}.

\noindent Ideally, to filter out the high \emph{SL} appearing in the
array pattern at $\theta=\theta_{GL}$ by also steering the main beam
along $\theta_{s}$, the $q$-th ($q=\left\{ \theta,\,\varphi\right\} $)
active element pattern of the $n$-th ($n=1,...,N$) radiating element
of the array should afford a maximum in correspondence with $\theta=\theta_{s}$,
while having a null along the \emph{GL} of the \emph{AF} (i.e., $\left|\mathcal{E}_{nq}\left(\theta,\,\varphi\right)\right|_{\theta=\theta_{GL};\,\varphi=0\,[\mathrm{deg}]}=0$)
\cite{Yepes 2020.b}-\cite{Ramaccia 2022}\cite{Ye 2019}. When dealing
with real-array elements, an effective strategy to approximate such
an ideal behavior is to mechanically tilt \emph{}each radiator along
the steering direction ($\theta_{0}=\theta_{s}$, $\theta_{0}$ being
the tilt angle) so that the radiation above the horizon is minimized
{[}i.e., $\left|\mathcal{E}_{nq}\left(\theta,\,\varphi\right)\right|_{\theta<90,\,\varphi=0\,[\mathrm{deg}]}$
$<$ $\left|\mathcal{E}_{nq}\left(\theta,\,\varphi\right)\right|_{\theta>90,\,\varphi=0\,[\mathrm{deg}]}$
$\le$ $\left|\mathcal{E}_{nq}\left(\theta,\,\varphi\right)\right|_{\theta=\theta_{0},\,\varphi=0\,[\mathrm{deg}]}$
- Fig. 2(\emph{b}){]}. Such an asymmetric \emph{angular selectivity}
at the array-element level enables the spatial filtering of the \emph{AF}
by strongly limiting the power radiated by the array towards $\theta_{GL}$
and, as a by-product, enhancing it along $\theta_{s}$ \cite{Yepes 2020.b}
{[}Fig. 2(\emph{b}){]}. However, the mechanical tilt of the array
elements is a limiting factor for the implementation of low-profile
structures. 

\noindent Otherwise, the same effects on the radiated \emph{FF} pattern
can be also yielded by still acting on (i.e., steering) the element
pattern, but without varying the orientation of the radiators to minimize
the array thickness. Towards this end, \emph{OSP}s \cite{Rajo-Iglesias 2002}-\cite{Rajo-Iglesias 2003}
are used by exploiting their standard working principle, but for a
purpose different from the standard one. Indeed, \emph{OSP}s are generally
designed either to improve the bandwidth of classical stacked patches
(i.e., with the top parasitic element aligned with respect to the
bottom excited element) \cite{Rajo-Iglesias 2002}-\cite{Luk 1993}
or to compensate the deviation of the pointing direction of probe-fed
patches by displacing the upper parasitic radiator in the opposite
direction to the feeding point \cite{Rajo-Iglesias 2003}. Differently,
they are here the technological recipe for building a low-profile
microstrip radiator with tilted radiation pattern. More in detail,
this paper considers a compact structure comprising $\mathcal{L}=3$
stacked dielectric layers of relative permittivity $\varepsilon_{\ell}$,
loss tangent $\tan\delta_{\ell}$, and thickness $t_{\ell}$ ($\ell=1,...,\mathcal{L}$)
(Fig. 3). The two bottom layers ($\ell=1,\,2$) implement a linearly-polarized
aperture-coupled patch where a rectangular slot of size $\left(L_{s}\times W_{s}\right)$
is etched in the ground plane on the top face of layer $\ell=1$ with
an offset $O_{s}$ along the $y$-axis {[}Fig. 4(\emph{a}){]}. Such
an aperture is fed by a microstrip line of length $L_{f}$ and width
$W_{f}$, which is terminated with an open-circuit stub and printed
on the bottom face of the same layer {[}Fig. 4(\emph{a}){]}. The driven
element is a rectangular patch of dimensions $\left(L_{p}\times W_{p}\right)$
on top of the layer $\ell=2$ {[}Fig. 4(\emph{b}){]}. An offset parasitic
patch is printed on the top layer of the antenna layout {[}i.e., $\ell=3$
- Fig. 3 and Fig. 4(\emph{c}){]} to steer the radiation pattern of
the radiator towards $\theta_{0}$ ($\theta_{0}>90$ {[}deg{]}). Such
a rectangular metallization, which is shifted by $\Delta_{z}$ along
the negative $z$-axis {[}Fig. 4(\emph{c}){]} and having dimensions
$\left(L_{d}\times W_{d}\right)$, acts as an offset {}``director''
of the electromagnetic (\emph{EM}) waves radiated by the underlying
patch that deviate from the broadside direction ($\theta=90$ {[}deg{]})
to generate an asymmetric element pattern with the desired spatial
filtering capability {[}Fig. 3(\emph{b}){]}.

\section{Co-Design Strategy \label{sec:Synthesis-Methodology}}

\noindent Once the materials and thicknesses $\left\{ \left(\varepsilon_{\ell},\,\tan\delta_{\ell},\, t_{\ell}\right);\,\ell=1,...,\mathcal{L}\right\} $
have been chosen%
\footnote{\noindent Among the off-the-shelf commercial products, the selection
is driven by the user needs in terms of cost, robustness, and potential
applications (e.g., terrestrial or space, environmental temperature/humidity).%
}, the design problem is recast to retrieve the optimal setup of the
$K$ geometric descriptors (i.e., $\underline{\chi}=\left\{ \chi_{k};\, k=1,...,K\right\} $,
$K=10$) of the \emph{OSP} (Fig. 4)\begin{equation}
\underline{\chi}=\left\{ L_{f},\, W_{f},\, L_{s},\, W_{s},\, O_{s},\, L_{p},\, W_{p},\, L_{d},\, W_{d},\,\Delta_{z}\right\} \label{DoFs}\end{equation}
to yield a suitable resonating behavior within the target frequency
band $B$ ($B\triangleq\left\{ f_{\min}\le f\le f_{\max}\right\} $)
as well as the desired spatial filtering properties when embedded
in a linear array of identical elements. 

\noindent It is worthwhile to notice that generally there are non-negligible
differences, in terms of both input impedance and element pattern,
between the \emph{EM} behavior of the embedded radiator and the stand-alone
one due to the unavoidable \emph{MC} effects. Moreover, the active
element pattern of each radiating element depends on its location
within the aperture \cite{Pozar 1994}. As a consequence, the synthesis
of $\underline{\chi}$ based on the analysis of the \emph{EM} features
of a single \emph{OSP} element radiating in free-space {[}Fig. 3(\emph{a}){]}
may lead to unsatisfactory solutions once embedded in a linear arrangement.

\noindent To yield a more robust and reliable solution, the set of
\emph{OSP} \emph{micro-scale} descriptors (\ref{DoFs}) is determined
through a co-design approach by directly optimizing the \emph{macro-scale}
\emph{EM} features of the corresponding finite-size array. In other
words, unlike forcing a proper impedance matching and a stand-alone
\emph{OSP} pattern towards $\theta_{0}=\theta_{s}$, the design of
the elementary radiator is carried out by requiring the embedded elements
with \emph{MC} to properly resonate in the target band, while lowering
the sidelobe level (\emph{SLL}) of the finite array when steering
the main beam along $\theta_{s}$. 

\noindent Towards this end, a global optimization is formulated by
defining the following cost function\begin{equation}
\Phi\left(\underline{\chi}\right)=\Phi_{EIM}\left(\underline{\chi}\right)+\Phi_{SLL}\left(\underline{\chi}\right).\label{eq:cost-function}\end{equation}
In (\ref{eq:cost-function}), $\Phi_{EIM}$ is the {}``embedded impedance
matching'' term, which enforces all the radiating elements of the
array to resonate in the frequency-band $B$, given by\begin{equation}
\Phi_{EIM}\left(\underline{\chi}\right)=\frac{1}{N\times B}{\displaystyle \sum_{n=1}^{N}}{\displaystyle \int_{f_{\min}}^{f_{\max}}\frac{\mathcal{R}\left\{ S_{nn}\left(\left.f\right|\underline{\chi}\right)-S^{th}\right\} }{\left|S^{th}\right|}df}\label{eq:S11-term}\end{equation}
where $S^{th}$ is a user-defined target threshold and $S_{nn}\left(\left.f\right|\underline{\chi}\right)$
is the reflection coefficient at the input port of the $n$-th ($n=1,...,N$)
array element, while $\mathcal{R}\left\{ \,.\,\right\} $ is the ramp
function (i.e., $\mathcal{R}\left\{ \xi\right\} =1$ if $\xi>0$,
$\mathcal{R}\left\{ \xi\right\} =0$ otherwise).

\noindent The other term in (\ref{eq:cost-function}) is devoted to
counteract the occurrence of the \emph{GL} (\ref{eq: GL}) in the
\emph{AF} of the array. It is defined as follows\begin{equation}
\Phi_{SLL}\left(\underline{\chi}\right)=\frac{1}{B}{\displaystyle \int_{f_{\min}}^{f_{\max}}\frac{1}{\left|SLL\left(\left.f\right|\underline{\chi}\right)\right|}df},\label{eq:SLL-term}\end{equation}
$SLL\left(\left.f\right|\underline{\chi}\right)$ being the \emph{SLL}
of the array when setting its excitations to steer the main beam towards
the user-defined down-looking angle $\theta=\theta_{s}$ (\ref{eq:excitations}).

\noindent Owing to both the high computational complexity of the synthesis
problem at hand and the need for a faithful prediction of the \emph{EM}
behavior of the real finite array to assess through (\ref{eq:cost-function})
the optimality of the antenna layout, the optimization {[}i.e., the
minimization of (\ref{eq:cost-function}){]} is carried within the
\emph{SbD} framework \cite{Massa 2021b}\cite{Salucci 2021}. More
specifically, the \emph{PSO-OK/C} implementation \cite{Massa 2021b}
of the \emph{SbD} paradigm is here exploited to remarkably speed up
the design process, while still keeping an accurate/reliable \emph{EM}
model of the array layout. It is based on the integration of a {}``\emph{Solution
Space Exploration}'' (\emph{SSE}) functional block, which leverages
the particle swarm optimization (\emph{PSO}) evolutionary operators
\cite{Rocca 2009}\cite{Goudos 2021}, with an accurate/fast surrogate
model (\emph{SM}) of (\ref{eq:cost-function}) relying on the Ordinary
Kriging (\emph{OK}) prediction technique \cite{Massa 2018b}\cite{Forrester 2008}\cite{Jones 1998}.
To complete the whole optimization process (i.e., the building of
the \emph{SM} and the iterative \emph{PSO}-based minimization of the
\emph{OK}-predicted cost function) within a fixed and limited amount
of time, the \emph{SM} is obtained by projecting the original solution
space of the design variables (\ref{DoFs}) within a reduced one to
minimize the size of the training set, thus reducing the number of
computationally-expensive full-wave simulations of (\ref{eq:cost-function}).

\noindent The co-design strategy can be summarized into the sequence
of the following procedural steps:

\begin{enumerate}
\item \emph{Input Phase -} Set the number of array elements, $N$, the inter-element
distance $d$, and the optimized down-looking elevation angle $\theta_{s}$.
Choose the dielectric materials and the thicknesses of the $\mathcal{L}$
layers composing the \emph{OSP} radiator. Set the desired \emph{EIM}
threshold $S^{th}$ (\ref{eq:S11-term});
\item \emph{SM} \emph{Initialization -} Sample the $K$-dimensional space
with the Latin hypercube sampling (\emph{LHS}) technique \cite{Massa 2021b}\cite{Salucci 2021}
to generate $B_{0}$ trial designs $\left\{ \underline{\chi}^{\left(b\right)};\, b=1,...,B_{0}\right\} $.
Simulate each $b$-th ($b=1,...,B_{0}$) design with a \emph{FW} solver
to compute its corresponding cost function, $\Phi\left(\underline{\chi}^{\left(b\right)}\right)$
($b=1,...,B_{0}$). Train an \emph{OK} model of the \emph{EM} behavior
of the \emph{OSP} radiator by using the reduced-dimensionality database
$\mathcal{D}_{0}=\left\{ \left[\aleph\left(\underline{\chi}^{\left(b\right)}\right),\,\Phi\left(\underline{\chi}^{\left(b\right)}\right)\right];\, b=1,...,B_{0}\right\} $($\aleph$
being the \emph{PLS} operator \cite{Massa 2021b}) to build a fast
\emph{SM}, $\widetilde{\Phi}_{0}\left(\underline{\chi}\right)$, of
the cost function $\Phi\left(\underline{\chi}\right)$ (\ref{eq:cost-function});
\item \emph{SSE Initialization} ($i=0$) - Initialize a swarm of $P$ particles
with random velocities $\mathcal{V}_{0}=\left\{ \underline{v}_{0}^{\left(p\right)};\, p=1,...,P\right\} $
and positions $\mathcal{P}_{0}=\left\{ \underline{\chi}_{0}^{\left(p\right)};\, p=1,...,P\right\} $
by setting\begin{equation}
\underline{\chi}_{0}^{\left(1\right)}=\arg\left\{ \min_{b=1,...,B_{0}}\left[\Phi\left(\underline{\chi}^{\left(b\right)}\right)\right]\right\} \label{eq:}\end{equation}
and\begin{equation}
\left\{ \begin{array}{c}
\underline{\chi}_{0}^{\left(p\right)}=\Psi\left\{ \left[\underline{\chi}^{\left(b\right)};\, b=1,...,B_{0}\right]\right\} \\
\underline{\chi}_{0}^{\left(p\right)}\ne\underline{\chi}_{0}^{\left(r\right)}\,\,\,\,\,\,\,\,\,\,\,\,\,\,\,\,\,\,\,\,\,\,\,\, r=1,...,p-1\end{array}\right.\label{eq:}\end{equation}
($p=2,...,P$), $\Psi\left\{ \,.\,\right\} $ being an operator randomly
picking an entry from an input set;
\item \emph{SSE Optimization} ($i=1,...,I$) - Iteratively evolve the swarm
positions by applying the \emph{PSO-OK/C} updating rules \cite{Massa 2021b}.
At each $i$-th ($i=1,...,I$) iteration select one trial solution
$\underline{\chi}_{i}^{\left(*\right)}$ on the basis of the \emph{OK}
confidence level and predict its EM behavior with a \emph{FW} solver
to compute the corresponding cost function value $\Phi\left(\underline{\chi}_{i}^{\left(*\right)}\right)$.
Add such a new training sample to the database, $\mathcal{D}_{i}\leftarrow\mathcal{D}_{i-1}\cup\left[\aleph\left(\underline{\chi}_{i}^{\left(*\right)}\right),\,\Phi\left(\underline{\chi}_{i}^{\left(*\right)}\right)\right]$,
and update the size of the training set, $B_{i}\leftarrow\left(B_{i-1}+1\right)$.
Retrain the \emph{OK} model with the updated database $\mathcal{D}_{i}=\left\{ \left[\aleph\left(\underline{\chi}^{\left(b\right)}\right),\,\Phi\left(\underline{\chi}^{\left(b\right)}\right)\right];\, b=1,...,B_{i}\right\} $
to update the \emph{SM}, $\widetilde{\Phi}_{i}\left(\underline{\chi}\right)\leftarrow\widetilde{\Phi}_{i-1}\left(\underline{\chi}\right)$
for adaptively enhancing its prediction accuracy within the attraction
basin of the global optimum of the cost function (\ref{eq:cost-function});
\item \emph{Output Phase} - Output the optimal layout by setting $\underline{\chi}^{\left(opt\right)}=\arg\left\{ \min_{b=1,...,B}\left[\Phi\left(\underline{\chi}^{\left(b\right)}\right)\right]\right\} $.
\end{enumerate}

\section{Numerical Assessment \label{sec:Numerical-Assessment}}

This section is aimed at assessing the effectiveness and the potentialities
of the proposed \emph{SFT} in synthesizing low-profile linear microstrip
arrays with wide-band spatial filtering capabilities.

\noindent The first numerical experiment is concerned with the design
of a linear \emph{PA} of $N=3$ horizontally-polarized (\emph{H-Pol})
\emph{OSP} elements, which are uniformly-spaced by $d=\lambda$, steered
along $\theta_{s}=110$ {[}deg{]} and operating within the frequency
range $B$ centered at $f_{c}=28$ {[}GHz{]} and ranging from $f_{\min}=26$
{[}GHz{]} up to $f_{\max}=30$ {[}GHz{]} ($\to$ $FBW=14.3\%$, $FBW$
being the fractional bandwidth), $S^{th}=-10$ {[}dB{]} being the
admissibility threshold for the reflection coefficient. By choosing
the \emph{Taconic TLY} {[}\emph{tm}{]} \emph{}as dielectric material
with $\varepsilon_{\ell}=2.2$, $\tan\delta_{\ell}=9\times10^{-4}$,
and $t_{\ell}=508$ {[}$\mu m${]} ($\ell=1,...,\mathcal{L}$), the
\emph{OSP} array layout has been forced to be low-profile with a total
thickness $T$ ($T\triangleq\sum_{\ell=1}^{\mathcal{L}}t_{\ell}$)
of $T=0.14\,\left[\lambda_{c}\right]$.

\noindent The co-design process has been carried out according to
the procedure in Sect. \ref{sec:Synthesis-Methodology} by setting
the control parameters according to the literature guidelines \cite{Massa 2021b}:
$P=10$ ($P$ - swarm size), $I=100$ ($I$ - number of iterations),
and $B_{0}=50$ ($B_{0}$ - size of the initial training set). Such
a setup allowed a time saving of $\Delta t=85\%$ \cite{Massa 2021b}
with respect to a standard \emph{PSO}-based optimization that exploits
only \emph{FW} simulations%
\footnote{\noindent The \emph{FW} Software Ansys \emph{HFSS} \cite{HFSS 2021}
has been used to infer the \emph{EM} behavior of the finite-size array.%
} for predicting the cost function values (\ref{eq:cost-function}).

\noindent Figure 5 shows the arising array layout (\emph{PA-OSP})
whose descriptors, $\underline{\chi}^{\left(opt\right)}$, are listed
in Tab. I. The array properly resonates in the target band $B$ as
one can infer from Fig. 6(\emph{a}) where the plot of the reflection
coefficient at the input port of each $n$-th ($n=1,...,N$) embedded
\emph{OSP} radiator is given. As a matter of fact, $S_{nn}\left(f\right)\leq-11.2$
{[}dB{]} ($n=1,...,N$) when $f\in B$ and $S_{nn}\left(f_{c}\right)\leq-23.3$
{[}dB{]} ($n=1,...,N$). Moreover, the \emph{PA-OSP} features a remarkable
spatial property as pointed out by the behavior of the normalized
power pattern at $f=f_{c}$ {[}Fig. 6(\emph{b}){]}. Indeed, the \emph{SLL}
value turns out to be significantly smaller than that of an equally-spaced
array of reference patches without offset directors (\emph{PA-RP}).
More in detail, the highest sidelobe appearing in the \emph{PA-RP}
pattern at $\theta_{SL}=48.85$ {[}deg{]}, which corresponds to the
\emph{GL} of the \emph{AF} {[}i.e., $\theta_{SL}=\theta_{GL}$ (\ref{eq: GL}){]},
is reduced by $\Delta SLL=9.49$ {[}dB{]} ($\Delta SLL\triangleq\left.SLL\right|_{PA-RP}-\left.SLL\right|_{PA-OSP}$)
from $\left.SLL\right|_{PA-RP}=-2.96$ {[}dB{]} down to $\left.SLL\right|_{PA-OSP}=-12.45$
{[}dB{]} {[}Fig. 6(\emph{b}) and Tab. II{]}. Furthermore, the comparison
of the \emph{3D} gain patterns of both the \emph{PA-OSP} {[}Fig. 7(\emph{a}){]}
and the \emph{PA-RP} {[}Fig. 7(\emph{b}){]} highlight that, besides
a \emph{SLL} reduction, there is also an enhancement of the array
gain in the steering direction of $\Delta G_{s}=0.97$ {[}dB{]} {[}$\Delta G_{s}\triangleq\left.G_{s}\right|_{PA-OSP}-\left.G_{s}\right|_{PA-RP}$
being $G_{s}=G\left(\theta=\theta_{s},\,\varphi=0\right)${]} (Tab.
II).

\noindent As expected, the filtering \emph{}features of the \emph{PA-OSP}
are the result of the angular selectivity of each elementary radiator
that makes up the array (Fig. 8). As anticipated in Sect. \ref{sec:Design-Concept-and-Principle},
this arises from a tilting of all the embedded element (\emph{EE})
patterns towards the array down-looking angular direction $\theta_{s}$,
which has been yielded in the co-design process without directly enforcing
it in the cost function (\ref{eq:cost-function}). Indeed, despite
some slight angular shifts of the tilt angle $\theta_{0}$ of the
different radiators caused by the \emph{MC} effects {[}i.e., $\theta_{0}^{\left(n\right)}\neq\theta_{0}^{\left(m\right)}$
($n,\, m=1,...,N$; $n\neq m$){]}, it turns out that $\theta_{0}^{\left(n\right)}\simeq\theta_{s}$
($n=1,...,N$) (Fig. 8), $G_{s}^{\left(n\right)}\geq7.9$ {[}dB{]}
($n=1,...,N$) being the gain along $\theta=\theta_{s}$ (Fig. 8).
Such a tilting allows one to filter out the \emph{GL} occurring in
the \emph{AF} thanks to the minimum of the radiation of the \emph{OSP}
radiators along $\theta=\theta_{GL}$ (Fig. 8), unlike the \emph{RP}
ones.

\noindent For completeness, Figure 9 shows the surface current distribution
induced on the \emph{PA-OSP} offset directors. On the one hand, the
plot indicates that such parasitic metallizations have been properly
excited, through proximity-coupling by the underlying driven patches,
so that the power is radiated towards the down-looking direction of
the array. On the other hand, as expected, the behavior of the induced
current complies with the fundamental $TM_{10}$ mode \cite{Perlmutter 1985}.

\noindent To point out the wide-band spatial filtering feature of
the \emph{PA-OSP}, Figure 10 plots the values of the \emph{SLL} and
of the gain within the operative band $B$. Despite the shift of the
angular position of the \emph{GL} when varying the frequency {[}$\theta_{GL}=42.70$
{[}deg{]} at $f=f_{\min}=26$ {[}GHz{]} - Fig. 11(\emph{a}); $\theta_{GL}=45.97$
{[}deg{]} at $f=27$ {[}GHz{]} - Fig. 11(\emph{b}); $\theta_{GL}=51.43$
{[}deg{]} at $f=29$ {[}GHz{]} - Fig. 11(\emph{c}); $\theta_{GL}=53.75$
{[}deg{]} at $f=f_{\max}=30$ {[}GHz{]} - Fig. 11(\emph{d}){]}, there
is always a reduction of the \emph{SLL} with respect to the \emph{PA-RP}
that amounts to $\left.\Delta SLL\right|_{f=f_{\min}}=4.75$ {[}dB{]}
($\left.\Delta SLL\right|_{f=f_{\max}}=7.51$ {[}dB{]}) at the lowest
(highest) frequency of $B$ (Fig. 10). The suppression of the sidelobes
of the pattern radiated by the \emph{PA-OSP} in the whole resonating
spectrum is made evident by the representative pattern plots in Fig.
11, which are very similar to that in Fig. 6(\emph{b}) for the central
frequency. Moreover, the \emph{PA-OSP} always yields a higher gain
than the \emph{PA-RP}, the improvement being equal to $\left.\Delta G_{s}\right|_{f=f_{\min}}=2.35$
{[}dB{]} ($\left.\Delta G_{s}\right|_{f=f_{\max}}=2.91$ {[}dB{]})
at the minimum (maximum) frequency (Fig. 10).

\noindent While the stand-alone optimization of the \emph{OSP} radiator
is unreliable for fulfilling the user-required \emph{EM} behavior
of the $N$-element array and its optimization in a large array is
almost unfeasible from the computational viewpoint because of the
use of a \emph{FW EM} simulator, it is interesting to understand how
effective is the \emph{OSP} layout synthesized for a small array when
arranged in a larger finite lattice of identical radiating elements.
To address such a question, two linear \emph{PA}-\emph{OSP}s of $N=5$
{[}Fig. 12(\emph{a}){]} and $N=10$ {[}Fig. 12(\emph{b}){]} elements
have been built using the same \emph{OSP} radiator of the previous
test case ($N=3$). The results of this analysis are summarized in
Fig. 12 and Tab. II. As it can be inferred, both the reduction of
the \emph{SLL} and the improvement of the array gain with respect
to the \emph{PA-RP} are of the same order in magnitude of the \emph{PA}
with $N=3$ elements. Indeed, it turns out that ($\left.\Delta SLL\right|_{N=5}=9.77$
{[}dB{]}; $\left.\Delta G_{s}\right|_{N=5}=1.0$ {[}dB{]}) and ($\left.\Delta SLL\right|_{N=10}=10.97$
{[}dB{]}; $\left.\Delta G_{s}\right|_{N=10}=1.1$ {[}dB{]}) when $N=5$
{[}Fig. 12(\emph{a}) - Tab. II{]} and $N=10$ {[}Fig. 12(\emph{b})
- Tab. II{]}, respectively.

\noindent Let us now focus on the largest aperture at hand (i.e.,
$N=10$). Figure 13 shows the tilting of all the \emph{EE-OSP} patterns
towards $\theta_{s}$ analogously to the $N=3$ case. Moreover, the
plot of the \emph{SLL} and of the gain versus the frequency in Fig.
14(\emph{a}) confirms both the wide-band filtering capabilities and
the higher gain of the \emph{OSP}-array in the whole working band.
It is also relevant to point out the robustness of the spatial filtering
capabilities of the \emph{PA-OSP} versus the scan angle, $\theta_{s}^{'}$,
as highlighted in Fig. 14(\emph{b}). Indeed, while the layout optimization
has been carried out by setting the down-tilt of the array at $\theta_{s}=110$
{[}deg{]}, the plots of the \emph{SLL} indicate that the \emph{PA-OSP}
overcomes the \emph{PA-RP} (i.e., $\left.SLL\right|_{PA-OSP}<\left.SLL\right|_{PA-RP}$)
in a wide portion of the scan range where $SLL<SLL_{th}$ ($SLL_{th}=0$
{[}dB{]}). As a matter of fact, the threshold value $SLL_{th}$, which
corresponds to the condition $G\left(\theta=\theta_{s}^{'}\right)=G\left(\theta=\theta_{GL}\right)$,
is reached by the \emph{PA-OSP} at a significantly greater value of
the scan angle than that of the \emph{PA-RP} {[}i.e., $\left.\theta_{s}^{'}\right|_{PA-OSP}^{th}=146$
{[}deg{]} versus $\left.\theta_{s}^{'}\right|_{PA-RP}^{th}=120$ {[}deg{]}
- Fig. 14(\emph{b}){]}. On the other hand, there is a slight degradation
of the \emph{PA-OSP} performance as compared to the \emph{PA-RP} one
for the values of $\theta_{s}^{'}$ close-to-the-horizon (i.e., $90$
{[}deg{]} $\le\theta_{s}^{'}\le$ $91.7$ {[}deg{]}) {[}Fig. 14(\emph{b}){]}
since the \emph{OSP} radiators have lower gains than the \emph{RP}
ones along broadside (i.e., $\left.G^{\left(n\right)}\left(\theta\right)\right|_{OSP}^{\theta=90\,[\mathrm{deg}]}\leq7.7$
{[}dB{]} vs. $\left.G^{\left(n\right)}\left(\theta\right)\right|_{RP}^{\theta=90\,[\mathrm{deg}]}\leq8.8$
{[}dB{]}, $n=1,...,N$ - Fig. 13). For illustrative purposes, the
\emph{FF} patterns at some representative values of the scan angle
$\theta_{s}^{'}$ {[}the plots at $\theta_{s}^{'}=\theta_{s}=110$
{[}deg{]} being in Fig. 12(\emph{b}){]} are reported in Fig. 15.

\noindent To further assess the generality of the proposed \emph{SFT},
the next test case deals with the synthesis of a $N=10$ elements
vertically-polarized (\emph{V-Pol}) \emph{PA-OSP}. Towards this end,
the co-design strategy (Sect. \ref{sec:Synthesis-Methodology}) has
been applied to an array of \emph{OSP} radiators rotated by $90$
{[}deg{]} with respect to the layout in Figs. 3-4 to realize still
a linear polarization, but orthogonal to that of the previous test
cases. The layout synthesized when $\theta_{s}=110$ {[}deg{]} is
shown in Fig. 16, while the values of its geometric descriptors are
reported in Tab. I. Once again, the \emph{PA-OSP} pattern is significantly
better than that of an equivalent-lattice \emph{PA-RP} with the same
polarization {[}Fig. 17(\emph{a}){]} since also quantitatively $\left.\Delta SLL\right|_{V-Pol}=11.63$
{[}dB{]} and $\left.\Delta G_{s}\right|_{V-Pol}=3.49$ {[}dB{]} (Tab.
II). Similarly to the \emph{H-Pol} case, such improvements are the
result of the tilting of the embedded elementary patterns towards
$\theta_{s}$ {[}Fig. 17(\emph{b}){]}.

\noindent Finally, a comparison between the \emph{V-Pol} \emph{PA-OSP}
in Fig. 16 and a \emph{PA} steered at the same direction ($\theta_{s}=110$
{[}deg{]}) with the same lattice/polarization, but composed by skewed
dipoles (\emph{PA-SD}) \cite{Yepes 2020.b}, has been carried out.
The \emph{FF} patterns in Fig. 18 show that the \emph{SLL} is almost
identical {[}i.e., $\left.SLL\right|_{PA-OSP}^{\theta_{s}=110\,[\mathrm{deg}]}=-12.85$
{[}dB{]} vs. $\left.SLL\right|_{PA-SD}^{\theta_{s}=110\,[\mathrm{deg}]}=-12.93$
{[}dB{]} - Figs. 18-19 and Tab. II{]}, but the gain of the \emph{PA-OSP}
is remarkably higher (i.e., $\left.G_{s}\right|_{PA-OSP}^{\theta_{s}=110\,[\mathrm{deg}]}=18.92$
{[}dB{]} vs. $\left.G_{s}\right|_{PA-SD}^{\theta_{s}=110\,[\mathrm{deg}]}=16.39$
{[}dB{]} - Fig. 19 and Tab. II). Moreover, we cannot forget that the
profile of the \emph{PA-OSP} is significantly lower (i.e., $T_{PA-OSP}=0.14\,\left[\lambda_{c}\right]$
vs. $T_{PA-SD}=0.7\,\left[\lambda_{c}\right]$ \cite{Yepes 2020.b}
- Fig. 1) since no mechanical rotations of the elementary radiators
are involved.

\noindent For completeness, the behavior of both the \emph{SLL} and
the gain versus $\theta_{s}^{'}$ is shown in Fig. 19. As it can be
inferred, the \emph{PA-OSP} yields contemporarily a similar \emph{SLL}
and a higher gain than the \emph{PA-SD} with the scanning range $103$
{[}deg{]}$\leq\theta_{s}^{'}\leq134$ {[}deg{]}, around the value
used in the \emph{PA-OSP} synthesis (i.e., $\theta_{s}=110$ {[}deg{]})
($\rightarrow$ $-7$ {[}deg{]} $\le\Delta\theta_{s}^{'}\le$ $24$
{[}deg{]}, $\Delta\theta_{s}^{'}\triangleq\theta_{s}^{'}-\theta_{s}$).

\section{Experimental Validation \label{sec:Experimental-Validation}}

To experimentally assess the effectiveness of the proposed \emph{SFT}
concept as well as the reliability of the co-design synthesis method,
an \emph{H-Pol} \emph{PA-OSP} prototype with $N=8$ elements operating
at $f_{c}=26$ {[}GHz{]} {[}$f\in\left[24,\,28\right]$ {[}GHz{]}
($\rightarrow FBW=15.4\%)${]} has been realized via printed circuit
board (\emph{PCB}) manufacturing (Fig. 20). The bottom layer ($\ell_{1}$)
of the \emph{PA-OSP} has been realized with a Rogers \emph{RO3003}
material ($\varepsilon_{1}=3.0$, $\tan\delta_{1}=0.0013$) of thickness
$t_{1}=130$ {[}$\mu m${]}, while the layers $\ell_{2}$ and $\ell_{3}$
have been implemented by using a Rogers/Duroid 5880 material ($\varepsilon_{2}=\varepsilon_{3}=2.2$,
$\tan\delta_{2}=\tan\delta_{3}=0.009$) with thickness $t_{2}=t_{3}=508$
{[}$\mu m${]} ($\rightarrow T=0.11$ {[}$\lambda_{c}${]}). A corporate
feeding network, able to feed the array elements with a fixed linear
phase shifting to steer the main beam towards $\theta_{s}=110$ {[}deg{]},
has been etched on the bottom face of the layer $\ell_{1}$ by cascading
2-ways power dividers and connecting them through 50 {[}$\Omega${]}
microstrip lines {[}Fig. 20(\emph{b}) and Fig. 20(\emph{d}){]}. For
comparison, a linear \emph{PA-RP} with the same number of array elements
has been designed and prototyped by using the same substrates and
beam-forming network (Fig. 21).

\noindent Figure 22 shows the magnitude of the reflection coefficient
measured at the input connector (\emph{RS PRO SMA 27G}) of both \emph{PA}s.
Both prototypes properly resonate in the target band (i.e., $\left|S_{11}\left(f\right)\right|_{PA-OSP}^{meas}\leq-11.6$
{[}dB{]} when $f\in B$ and $\left|S_{11}\left(f_{c}\right)\right|_{PA-OSP}^{meas}=-12.9$
{[}dB{]} %
\footnote{\noindent The deviations of the measured $S_{11}$ values from the
simulated ones (Fig. 22) are due to several non-idealities in the
manufacturing process as well as to the presence of spurious reflections
caused by the soldered connector {[}Figs. 20(\emph{c})-20(\emph{d})
and Fig. 21(\emph{b}){]}. %
}.

\noindent Concerning the radiation features, Figure 23(\emph{a}) shows
the plots of the measured co-polar patterns at $f=f_{c}$. One can
notice that the undesired radiation towards $\theta_{GL}$ is considerably
limited, the level of the sidelobes being reduced from $\left.SLL\right|_{PA-RP}^{meas}=-2.3$
{[}dB{]} down to $\left.SLL\right|_{PA-OSP}^{meas}=-11.6$ {[}dB{]}
($\rightarrow$$\Delta SLL^{meas}=9.3$ {[}dB{]}) in close agreement
with the simulations (i.e., $\Delta SLL^{sim}=9.6$ {[}dB{]}).

\noindent Finally, Figure 23(\emph{b}) gives a proof of the polarization
purity of the array radiation by showing that the measured cross-polar
(\emph{CX}) level is $\left.CX\left(\theta\right)\right|_{PA-OSP}^{meas}\leq-25.1$
{[}dB{]} within the range $\theta\in\left[0,\,180\right]$ {[}deg{]},
while $\left.CX\left(\theta\right)\right|_{PA-RP}^{meas}\leq-24.2$
{[}dB{]}.

\section{\noindent Conclusions \label{sec:Conclusions}}

\noindent An innovative \emph{SFT} for implementing low-profile linear
microstrip \emph{PA}s with inter-element spacing larger than half-wavelength
has been presented. The proposed architecture exploits \emph{OSP}
radiators to implement an angular selectivity suitable for suppressing
the undesired \emph{PA} radiation caused by the insurgence of \emph{GL}s
in the corresponding \emph{AF}. To provide a reliable and robust design,
which takes into account the \emph{MC} effects arising in real/finite-size
arrays, a \emph{co-design} approach has been introduced where the
\emph{micro-scale} descriptors of the elementary \emph{OSP} radiator
are optimized to fulfill the requirements on the \emph{macro-scale}
\emph{EM} performance of the corresponding \emph{PA}.

\noindent The main outcomes from the numerical and experimental assessment
have been the following ones:

\begin{itemize}
\item the proposed \emph{SFT} yields a remarkable control of the \emph{SLL}
of the \emph{PA} over a wide-band (e.g., $FBW=14.3\%$); 
\item the \emph{OSP-PA} still guarantees a non-negligible suppression of
the \emph{GL}s also when steering the beam towards down-looking directions
different from the optimized one;
\item it is possible (i.e., the performance are still acceptable) to build
linear arrays of different size (i.e., a different number of array
elements) without the need for re-optimizing the layout of the \emph{OSP}
radiator for each \emph{PA} arrangement; 
\item both the \emph{PA} architecture and the co-design method can be seamlessly
used to synthesize \emph{H-Pol} as well as \emph{V-Pol} linear arrays;
\item the \emph{OSP-PA} solution positively compares with a leading-edge
state-of-the-art solution \cite{Yepes 2020.b} in terms of spatial
filtering capabilities and gain values, while exhibiting a lower profile;
\item the \emph{FW}-simulated performance and the \emph{EM} behavior of
the proposed \emph{OSP-PA} architecture are experimentally confirmed
by a \emph{PCB}-manufactured \emph{}prototype operating in the \emph{mm}-wave
band $24$ {[}GHz{]} \emph{$\le f\le$} $28$ {[}GHz{]}.
\end{itemize}
\noindent Future works, beyond the scope of this paper, will be aimed
at extending the proposed \emph{SFT} to both planar arrangements and
dual-polarizations (e.g., slant-45) operation \cite{Ramaccia 2022}.

\section*{\noindent Acknowledgements}

\noindent This work benefited from the networking activities carried
out within the Project Project {}``Hub Life Science - Advanced Diagnosis
(HLS-AD), PNRR PNC-E3-2022-23683266 PNC-HLS-DA, INNOVA'' funded by
the Italian Ministry of Health under the National Complementary Plan
Innovative Health Ecosystem (CUP: E63C22003780001 - Unique Investment
Code: PNC-E.3), the Project {}``AURORA - Smart Materials for Ubiquitous
Energy Harvesting, Storage, and Delivery in Next Generation Sustainable
Environments'' funded by the Italian Ministry for Universities and
Research within the PRIN-PNRR 2022 Program (CUP: E53D23014760001),
the Project {}``Telecommunications of the Future'' (PE00000001 -
program {}``RESTART'', Structural Project 6GWINET), funded by European
Union under the Italian National Recovery and Resilience Plan (NRRP)
of NextGenerationEU (CUP: D43C22003080001), and the Project DICAM-EXC
(Grant L232/2016) funded by the Italian Ministry of Education, Universities
and Research (MUR) within the {}``Departments of Excellence 2023-2027''
Program (CUP: E63C22003880001). The authors are very grateful to Prof.
R. Azaro (EMC s.r.l., Via Greto di Cornigliano, 6/r, 16152 GENOVA
- emc.info@emclab.it) for the prototyping and the experimental measurements.
A. Massa wishes to thank E. Vico for her never-ending inspiration,
support, guidance, and help.

\newpage
\section*{FIGURE CAPTIONS}

\begin{itemize}
\item \textbf{Figure 1.} Linear array layouts.
\item \textbf{Figure 2.} Pictorial sketch of a linear array of $d$-spaced
($d>\frac{\lambda}{2}$) microstrip (\emph{a}) patches and (\emph{b})
offset stacked patches. 
\item \textbf{Figure 3.} Layout of (\emph{a}) the \emph{OSP} radiator and
(\emph{b}) the uniform linear array of $N$ \emph{OSP} radiators.
\item \textbf{Figure 4.} Top view of the $\ell$-th layer of the \emph{OSP}
radiator: $\ell=1$, (\emph{b}) $\ell=2$ , and (\emph{c}) $\ell=3$.
\item \textbf{Figure 5.} \emph{Numerical Assessment} (\emph{H-Pol}, $N=3$,
$\theta_{s}=110$ {[}deg{]}) - \emph{CAD} model of the synthesized
\emph{PA}-\emph{OSP}.
\item \textbf{Figure 6.} \emph{Numerical Assessment} (\emph{H-Pol}, $N=3$,
$\theta_{s}=110$ {[}deg{]}) - Plots of (\emph{a}) the reflection
coefficient at the input port of each $n$-th ($n=1,...,N$) embedded
\emph{OSP} radiator versus the frequency and (\emph{b}) the normalized
power pattern at $f=f_{c}$ along with the \emph{AF} and the \emph{EE}
pattern of the central element ($n=2$) of both the \emph{PA-RP} and
the \emph{PA-OSP} in the elevation cut ($\varphi=0$ {[}deg{]}).
\item \textbf{Figure 7.} \emph{Numerical Assessment} (\emph{H-Pol}, $N=3$,
$\theta_{s}=110$ {[}deg{]}, $f=f_{c}=28$ {[}GHz{]}) - Simulated
\emph{3D} gain pattern of (\emph{a}) the \emph{PA-OSP} and (\emph{b})
the \emph{PA-RP} arrays.
\item \textbf{Figure 8.} \emph{Numerical Assessment} (\emph{H-Pol}, $N=3$,
$\theta_{s}=110$ {[}deg{]}, $f=f_{c}=28$ {[}GHz{]}) - Simulated
gain patterns of the $N$ elements of both the \emph{PA-RP} and the
\emph{PA-OSP}.
\item \textbf{Figure 9.} \emph{Numerical Assessment} (\emph{H-Pol}, $N=3$,
$\theta_{s}=110$ {[}deg{]}, $f=f_{c}=28$ {[}GHz{]}) - Simulated
distribution of the surface current excited on the offset directors
of the \emph{PA-OSP} in Fig. 5.
\item \textbf{Figure 10.} \emph{Numerical Assessment} (\emph{H-Pol}, $N=3$,
$\theta_{s}=110$ {[}deg{]}) - Behavior of the \emph{SLL} and of the
gain ($G_{s}$) versus the frequency $f$ for both the \emph{PA-RP}
and the \emph{PA-OSP}.
\item \textbf{Figure 11.} \emph{Numerical Assessment} (\emph{H-Pol}, $N=3$,
$\theta_{s}=110$ {[}deg{]}) - Plot in the elevation cut ($\varphi=0$
{[}deg{]}) of the normalized power pattern along with the \emph{AF}
and the \emph{EE} pattern of the central element ($n=2$) of both
the \emph{PA-RP} and the \emph{PA-OSP} at (\emph{a}) $f=f_{\min}=26$
{[}GHz{]}, (\emph{b}) $f=27$ {[}GHz{]}, (\emph{c}) $f=29$ {[}GHz{]},
and (\emph{d}) $f=f_{\max}=30$ {[}GHz{]}.
\item \textbf{Figure 12.} \emph{Numerical Assessment} (\emph{H-Pol}, $\theta_{s}=110$
{[}deg{]}, $f=f_{c}=28$ {[}GHz{]}) - Plot in the elevation cut ($\varphi=0$
{[}deg{]}) of the normalized power pattern along with the \emph{AF}
and the \emph{EE} pattern of the central element of both the \emph{PA-RP}
and the \emph{PA-OSP} with (\emph{a}) $N=5$ and (\emph{b}) $N=10$
elements.
\item \textbf{Figure 13.} \emph{Numerical Assessment} (\emph{H-Pol}, $N=10$,
$\theta_{s}=110$ {[}deg{]}, $f=f_{c}=28$ {[}GHz{]}) - Simulated
gain patterns of the $N$ elements of both the \emph{PA-RP} and the
\emph{PA-OSP}.
\item \textbf{Figure 14.} \emph{Numerical Assessment} (\emph{H-Pol}, $N=10$,
$\theta_{s}=110$ {[}deg{]}) - Behavior of the \emph{SLL} and of the
gain ($G_{s}$) for both the \emph{PA-RP} and the \emph{PA-OSP} versus
(\emph{a}) the frequency $f$ and (\emph{b}) the scan angle $\theta_{s}^{'}$.
\item \textbf{Figure 15.} \emph{Numerical Assessment} (\emph{H-Pol}, $N=10$,
$f=f_{c}=28$ {[}GHz{]}, $\theta_{s}=110$ {[}deg{]}) - Plot in the
elevation cut ($\varphi=0$ {[}deg{]}) of the normalized power pattern
along with the \emph{AF} and the \emph{EE} pattern of the central
element of both the \emph{PA-RP} and the \emph{PA-OSP} when steering
the main beam towards (\emph{a}) $\theta_{s}^{'}=90$ {[}deg{]} ($\rightarrow\theta_{GL}=0.0$
{[}deg{]}), (\emph{b}) $\theta_{s}^{'}=100$ {[}deg{]} ($\rightarrow$
$\theta_{GL}=34.3$ {[}deg{]}), and (\emph{c}) $\theta_{s}^{'}=140$
{[}deg{]} ($\rightarrow$ $\theta_{GL}=76.5$ {[}deg{]}).
\item \textbf{Figure 16.} \emph{Numerical Assessment} (\emph{V-Pol}, $N=10$,
$\theta_{s}=110$ {[}deg{]}) - \emph{CAD} model of the synthesized
\emph{PA}-\emph{OSP}.
\item \textbf{Figure 17.} \emph{Numerical Assessment} (\emph{V-Pol}, $N=10$,
$\theta_{s}=110$ {[}deg{]}, $f=f_{c}=28$ {[}GHz{]}) - Plot in the
elevation cut ($\varphi=0$ {[}deg{]}) of (\emph{a}) the normalized
power pattern along with the \emph{AF} and the \emph{EE} pattern of
the central element of both the \emph{PA-RP} and the \emph{PA-OSP}
and (\emph{b}) the gain patterns of the $N$ elements of the \emph{PA-OSP}.
\item \textbf{Figure 18.} \emph{Numerical Assessment} (\emph{V-Pol}, $N=10$,
$\theta_{s}=110$ {[}deg{]}, $f=f_{c}=28$ {[}GHz{]}) - Plot in the
elevation cut ($\varphi=0$ {[}deg{]}) of the normalized power pattern
along with the \emph{AF} and the \emph{EE} pattern of the central
element of both the \emph{PA-SD} \cite{Yepes 2020.b} and the \emph{PA-OSP}.
\item \textbf{Figure 19.} \emph{Numerical Assessment} (\emph{V-Pol}, $N=10$,
$\theta_{s}=110$ {[}deg{]}, $f=f_{c}=28$ {[}GHz{]}) - Behavior of
the \emph{SLL} and of the gain ($G_{s}$) versus the scan angle $\theta_{s}^{'}$
for the \emph{PA-SD} \cite{Yepes 2020.b}, the \emph{PA-RP}, and the
\emph{PA-OSP}. 
\item \textbf{Figure 20.} \emph{Experimental Assessment} (\emph{H-Pol},
$N=8$, $\theta_{s}=110$ {[}deg{]}) - Pictures of (\emph{a})(\emph{c})
the top and (\emph{b})(\emph{d}) the bottom views of the (\emph{a})(\emph{b})
\emph{FW}-simulated model and (\emph{c})(\emph{d}) the \emph{PCB}-manufactured
prototype of the \emph{PA-OSP}.
\item \textbf{Figure 21.} \emph{Experimental Assessment} (\emph{H-Pol},
$N=8$, $\theta_{s}=110$ {[}deg{]}, $f=f_{c}=26$ {[}GHz{]}) - Pictures
of the top view of the (\emph{a}) \emph{FW}-simulated model and (\emph{b})
the \emph{PCB}-manufactured prototype of the \emph{PA-RP}.
\item \textbf{Figure 22.} \emph{Experimental Assessment} (\emph{H-Pol},
$N=8$, $\theta_{s}=110$ {[}deg{]}, $f\in\left[22,\,30\right]$ {[}GHz{]})
- Simulated and measured values of the reflection coefficient at the
input port of both the \emph{PA-OSP} and the \emph{PA-RP}.
\item \textbf{Figure 23.} \emph{Experimental Assessment} (\emph{H-Pol},
$N=8$, $\theta_{s}=110$ {[}deg{]}, $f=f_{c}=26$ {[}GHz{]}) - Simulated
and measured (\emph{a}) co-polar and (\emph{b}) cross-polar patterns
radiated by the \emph{PA-OSP} and the \emph{PA-RP}.
\end{itemize}

\section*{TABLE CAPTIONS}

\begin{itemize}
\item \textbf{Table I.} \emph{Numerical Assessment} ($d=\lambda$, $\left[f_{\min},\, f_{\max}\right]=\left[26,\,30\right]$
{[}GHz{]}, $\theta_{s}=110$ {[}deg{]}) - Optimized values of the
\emph{OSP} descriptors (\ref{DoFs}).
\item \textbf{Table II.} \emph{Numerical Assessment} ($d=\lambda$, $\left[f_{\min},\, f_{\max}\right]=\left[26,\,30\right]$
{[}GHz{]}, $\theta_{s}=110$ {[}deg{]}, $f=f_{c}=28$ {[}GHz{]}) -
\emph{SLL} and gain values.
\end{itemize}
\newpage
\begin{center}~\vfill\end{center}

\begin{center}\includegraphics[%
  width=0.95\columnwidth]{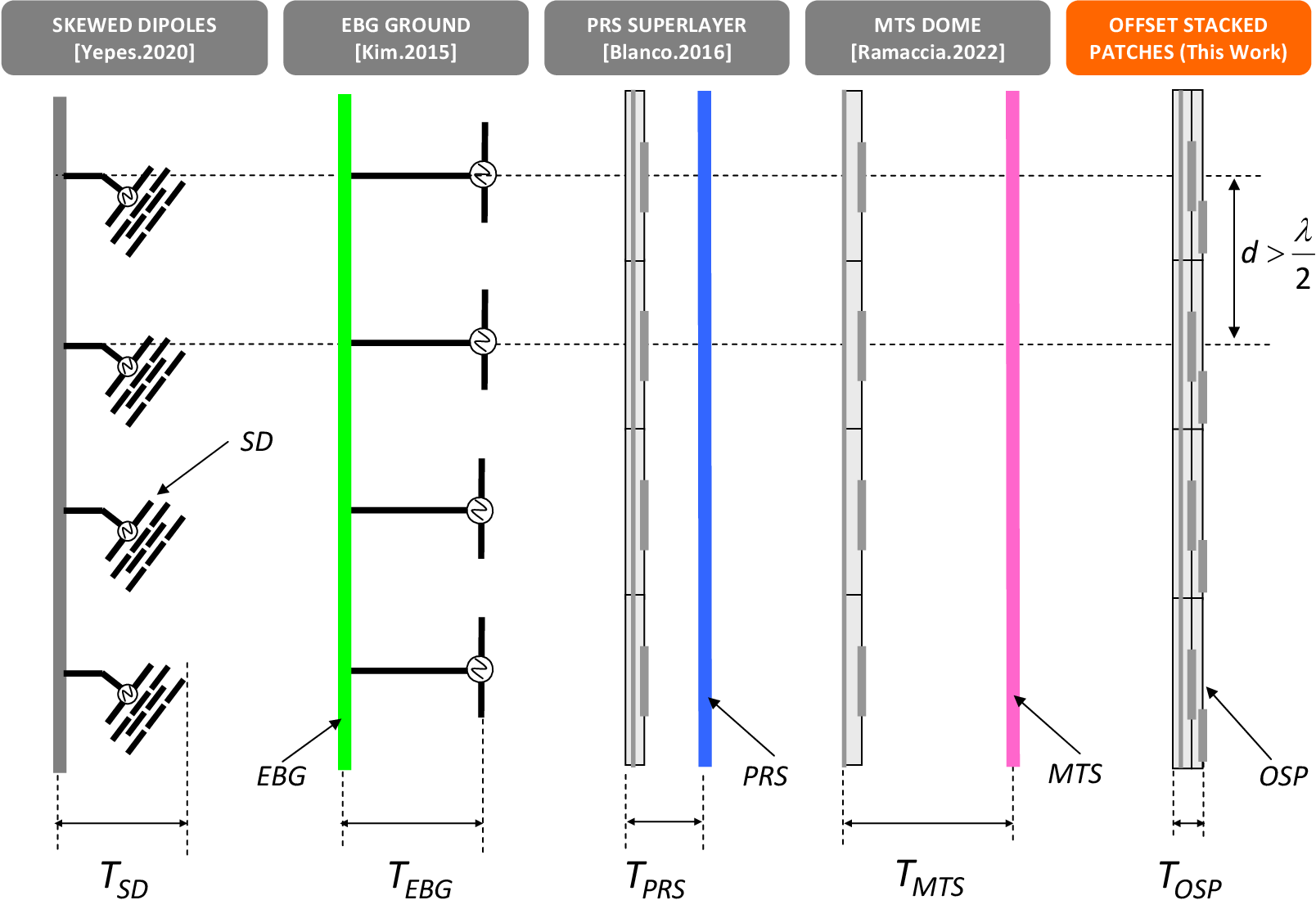}\end{center}

\begin{center}~\vfill\end{center}

\begin{center}\textbf{Fig. 1 - A. Benoni et} \textbf{\emph{al.}}\textbf{,}
\textbf{\emph{{}``}}Co-Design of Low-Profile Linear Microstrip Arrays
...''\end{center}

\newpage
\begin{center}~\vfill\end{center}

\begin{center}\begin{tabular}{ccc}
\multicolumn{3}{c}{\includegraphics[%
  width=0.75\columnwidth]{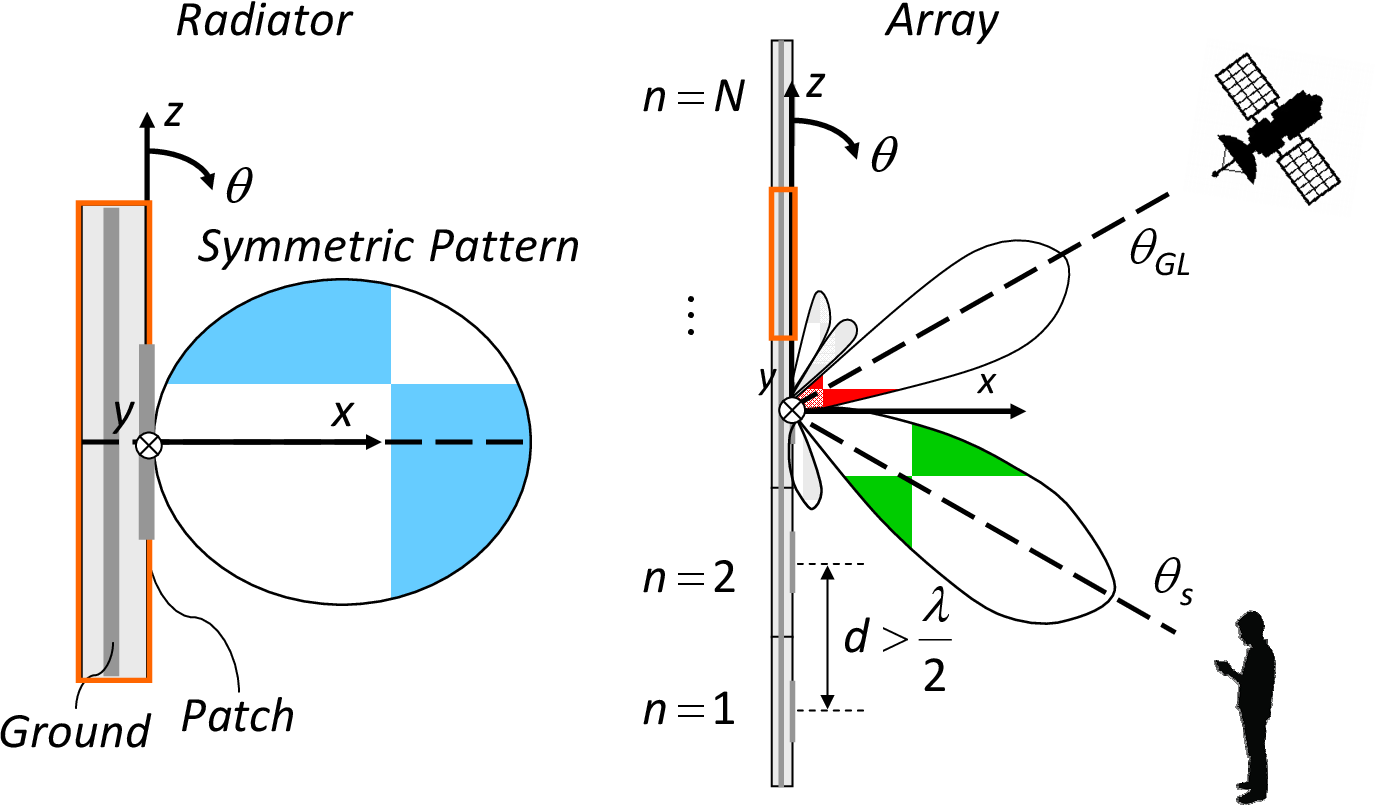}}\tabularnewline
\multicolumn{3}{c}{(\emph{a})}\tabularnewline
\multicolumn{3}{c}{}\tabularnewline
\multicolumn{3}{c}{\includegraphics[%
  width=0.75\columnwidth]{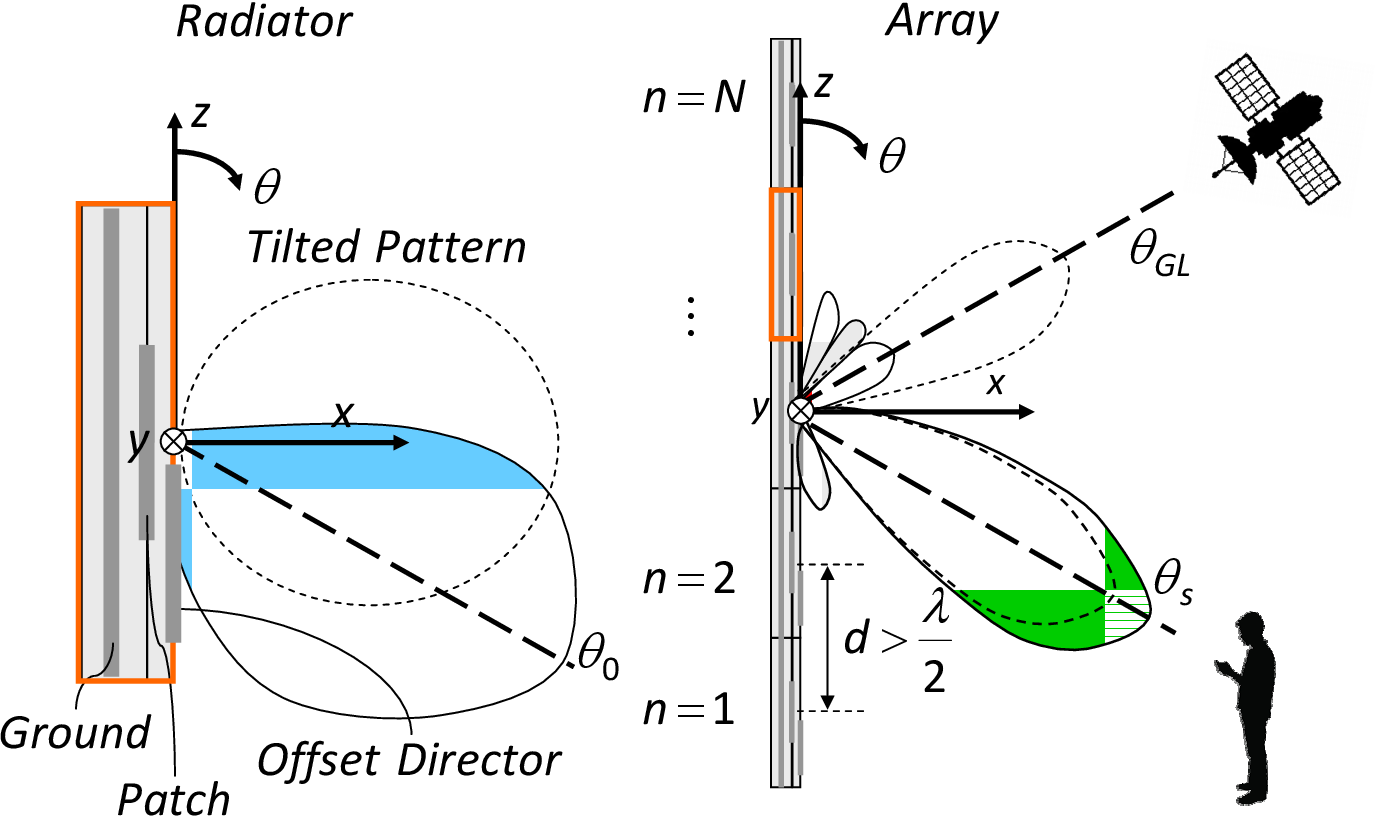}}\tabularnewline
\multicolumn{3}{c}{(\emph{b})}\tabularnewline
\end{tabular}\end{center}

\begin{center}~\vfill\end{center}

\begin{center}\textbf{Fig. 2 - A. Benoni et} \textbf{\emph{al.}}\textbf{,}
\textbf{\emph{{}``}}Co-Design of Low-Profile Linear Microstrip Arrays
...''\end{center}

\newpage
\begin{center}~\vfill\end{center}

\begin{center}\begin{tabular}{c}
\includegraphics[%
  width=0.90\columnwidth]{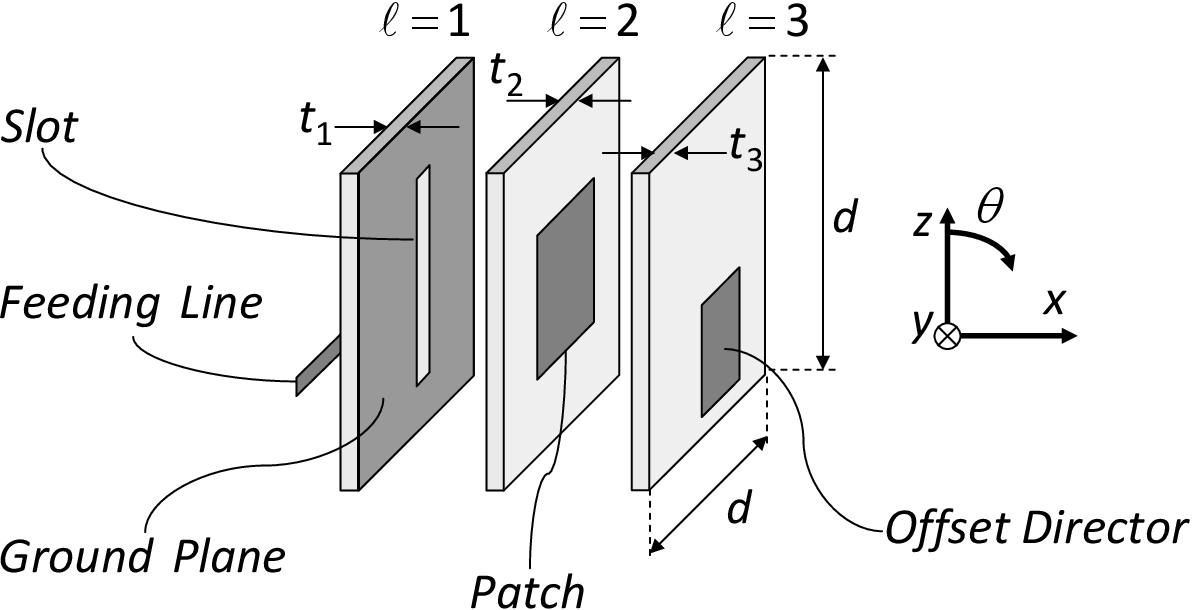}\tabularnewline
(\emph{a})\tabularnewline
\tabularnewline
\includegraphics[%
  width=0.40\columnwidth]{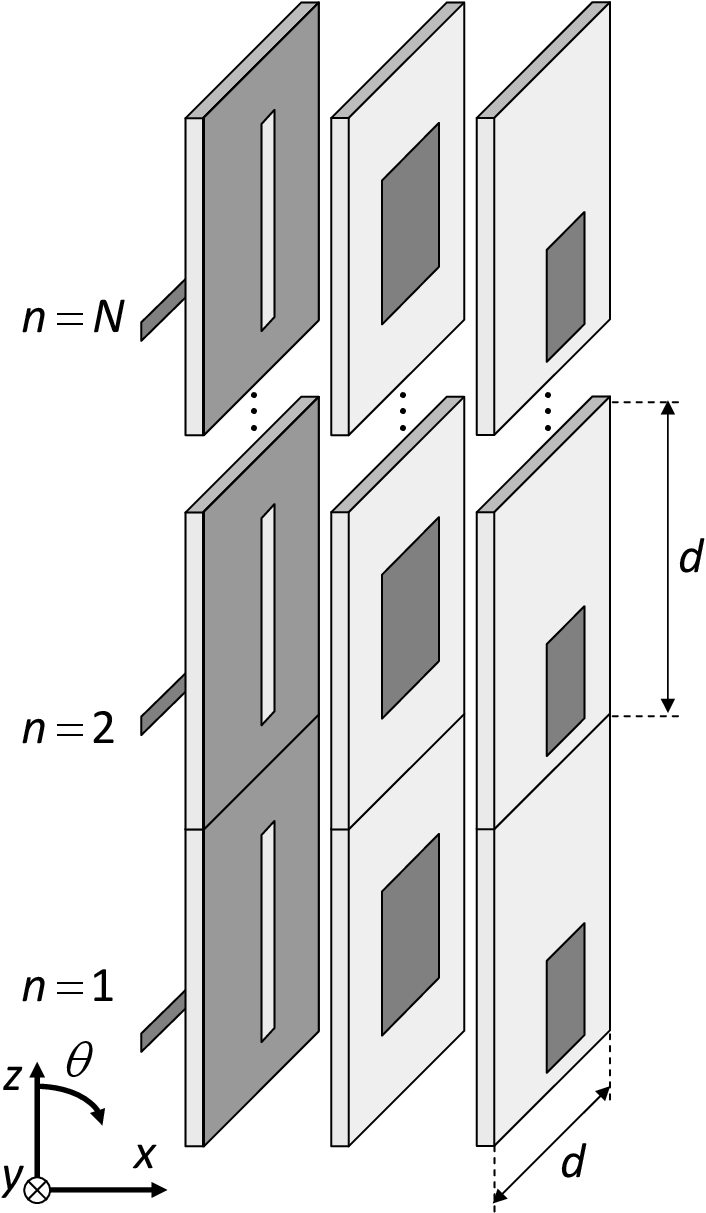}\tabularnewline
(\emph{b})\tabularnewline
\end{tabular}\end{center}

\begin{center}~\vfill\end{center}

\begin{center}\textbf{Fig. 3 - A. Benoni et} \textbf{\emph{al.}}\textbf{,}
\textbf{\emph{{}``}}Co-Design of Low-Profile Linear Microstrip Arrays
...''\end{center}

\newpage
\begin{center}~\vfill\end{center}

\begin{center}\begin{tabular}{cc}
\includegraphics[%
  width=0.45\columnwidth]{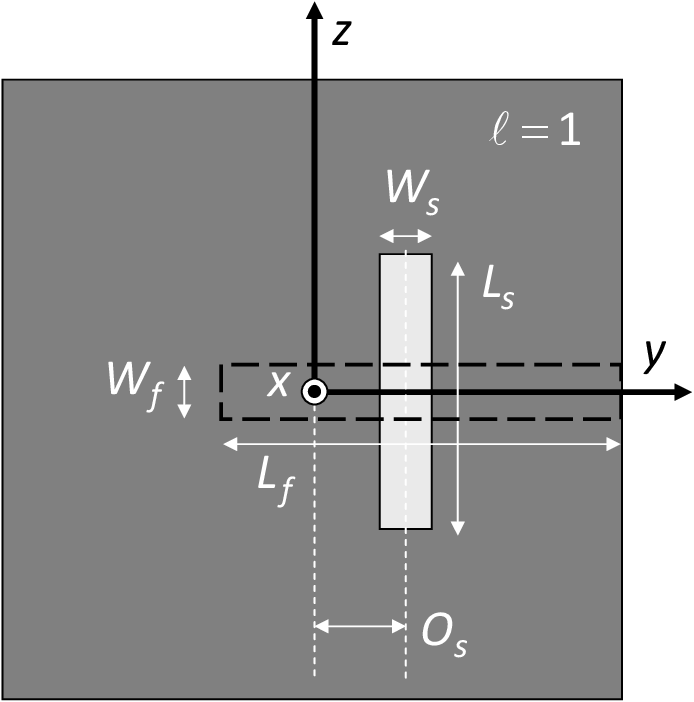}&
\includegraphics[%
  width=0.45\columnwidth]{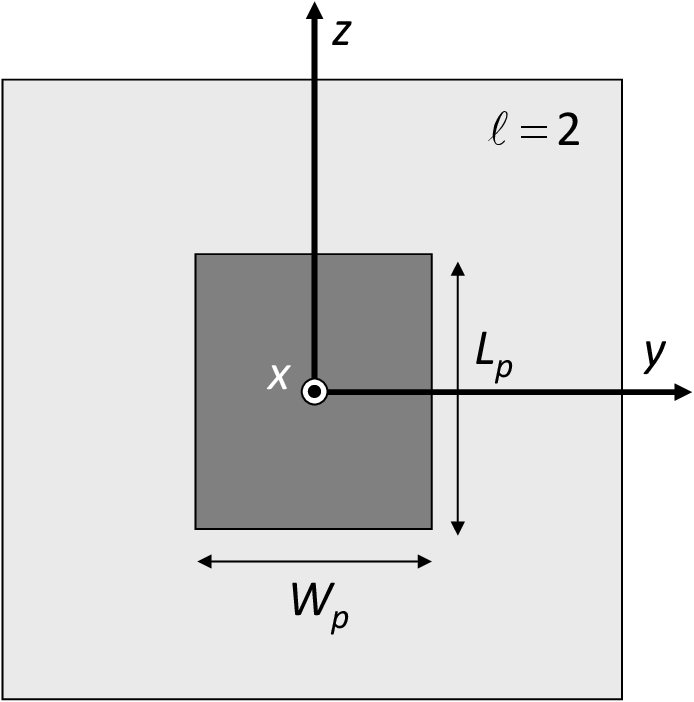}\tabularnewline
(\emph{a})&
(\emph{b})\tabularnewline
\multicolumn{2}{c}{}\tabularnewline
\multicolumn{2}{c}{\includegraphics[%
  width=0.45\columnwidth]{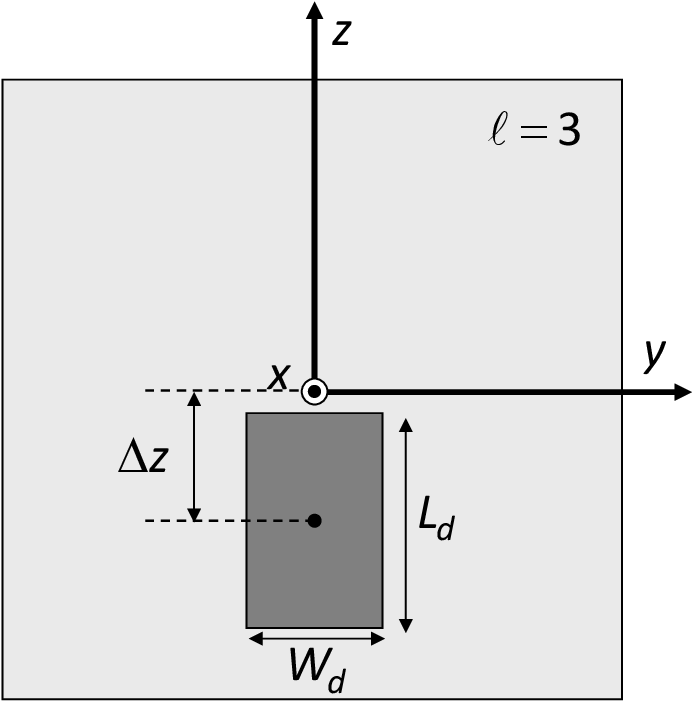}}\tabularnewline
\multicolumn{2}{c}{(\emph{c})}\tabularnewline
\end{tabular}\end{center}

\begin{center}~\vfill\end{center}

\begin{center}\textbf{Fig. 4 - A. Benoni et} \textbf{\emph{al.}}\textbf{,}
\textbf{\emph{{}``}}Co-Design of Low-Profile Linear Microstrip Arrays
...''\end{center}

\newpage
\begin{center}~\vfill\end{center}

\begin{center}\includegraphics[%
  width=0.65\columnwidth]{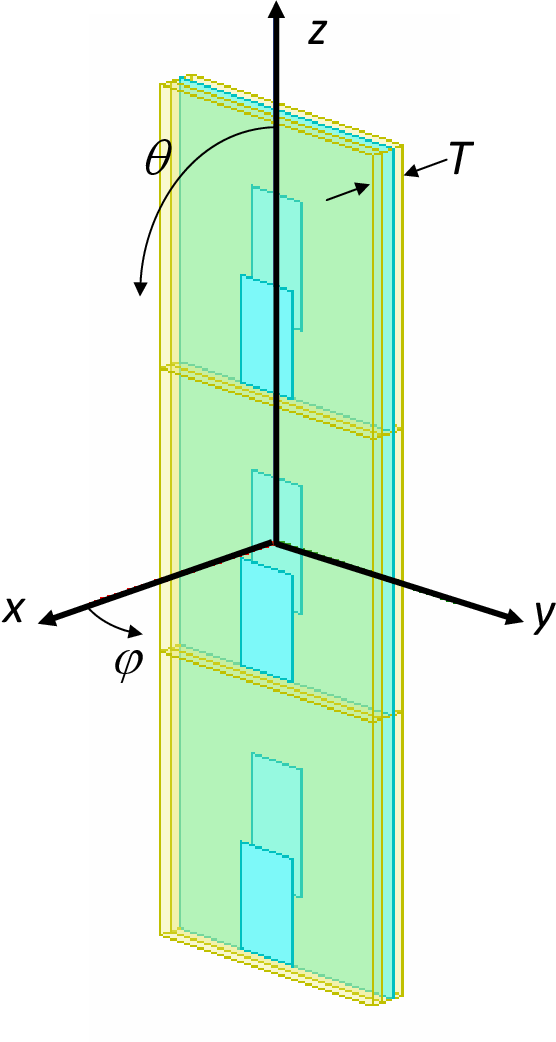}\end{center}

\begin{center}~\vfill\end{center}

\begin{center}\textbf{Fig. 5 - A. Benoni et} \textbf{\emph{al.}}\textbf{,}
\textbf{\emph{{}``}}Co-Design of Low-Profile Linear Microstrip Arrays
...''\end{center}

\newpage
\begin{center}~\vfill\end{center}

\begin{center}\begin{tabular}{c}
\includegraphics[%
  width=0.80\columnwidth]{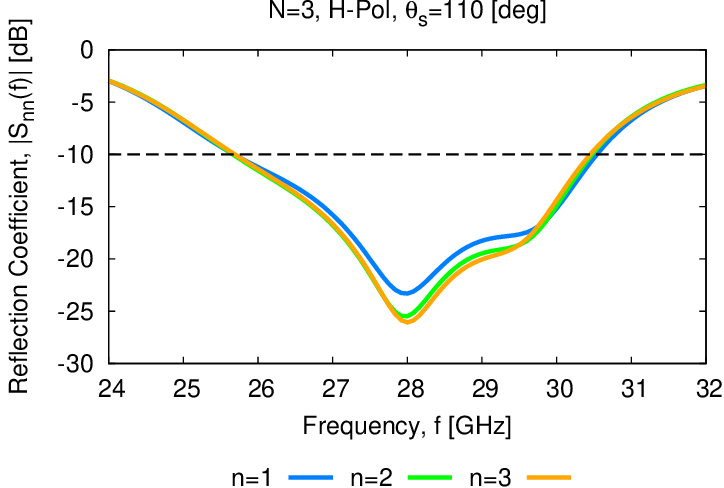}\tabularnewline
(\emph{a})\tabularnewline
\tabularnewline
\includegraphics[%
  width=0.80\columnwidth]{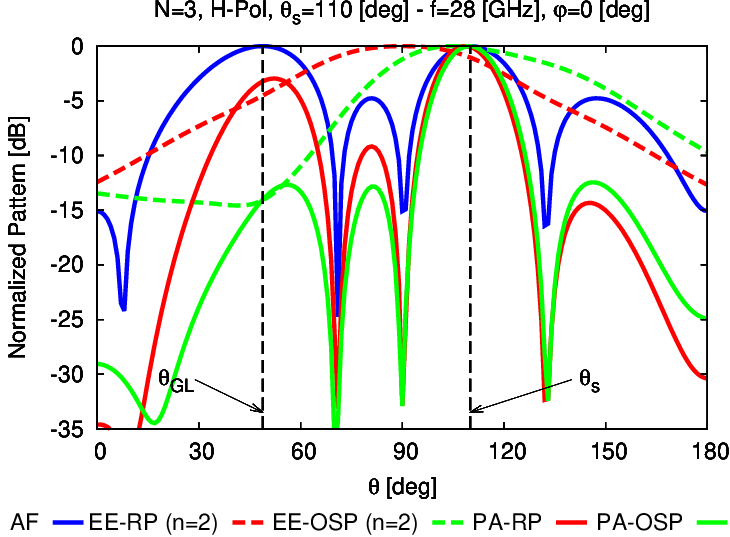}\tabularnewline
(\emph{b})\tabularnewline
\end{tabular}\end{center}

\begin{center}~\vfill\end{center}

\begin{center}\textbf{Fig. 6 - A. Benoni et} \textbf{\emph{al.}}\textbf{,}
\textbf{\emph{{}``}}Co-Design of Low-Profile Linear Microstrip Arrays
...''\end{center}

\newpage
\begin{center}~\vfill\end{center}

\begin{center}\begin{tabular}{c}
\includegraphics[%
  width=0.65\columnwidth]{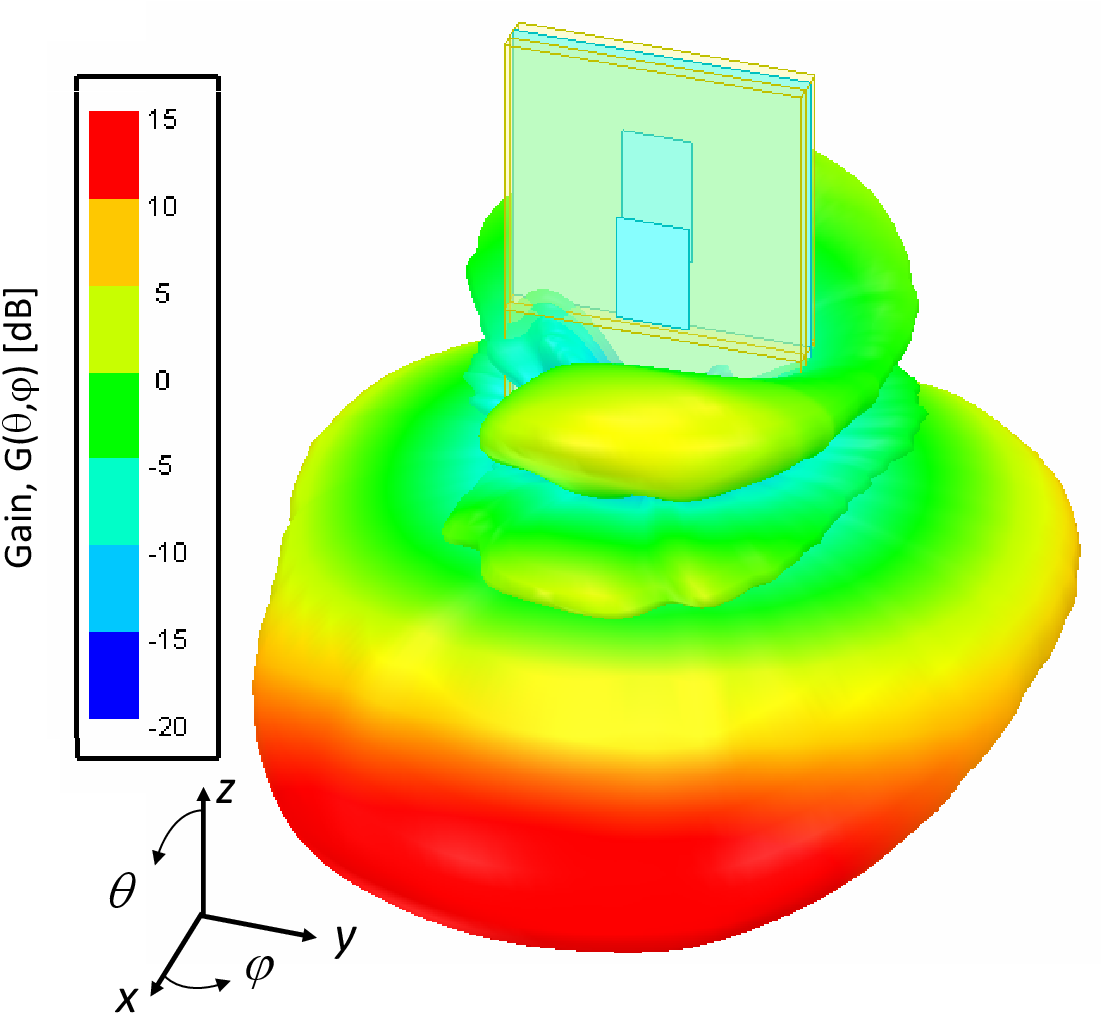}\tabularnewline
(\emph{a})\tabularnewline
\tabularnewline
\includegraphics[%
  width=0.65\columnwidth]{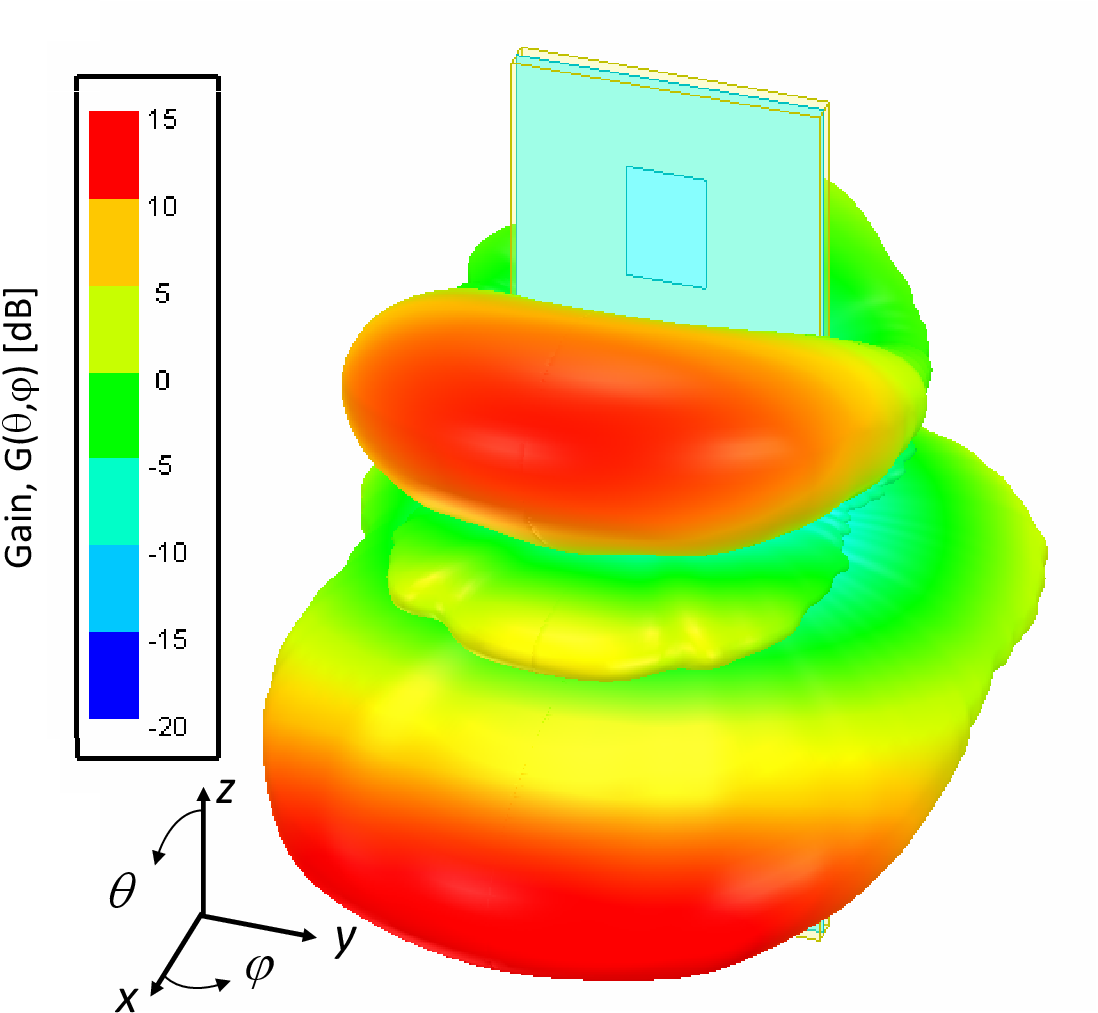}\tabularnewline
(\emph{b})\tabularnewline
\end{tabular}\end{center}

\begin{center}~\vfill\end{center}

\begin{center}\textbf{Fig. 7 - A. Benoni et} \textbf{\emph{al.}}\textbf{,}
\textbf{\emph{{}``}}Co-Design of Low-Profile Linear Microstrip Arrays
...''\end{center}

\newpage
\begin{center}~\vfill\end{center}

\begin{center}\includegraphics[%
  width=0.90\columnwidth]{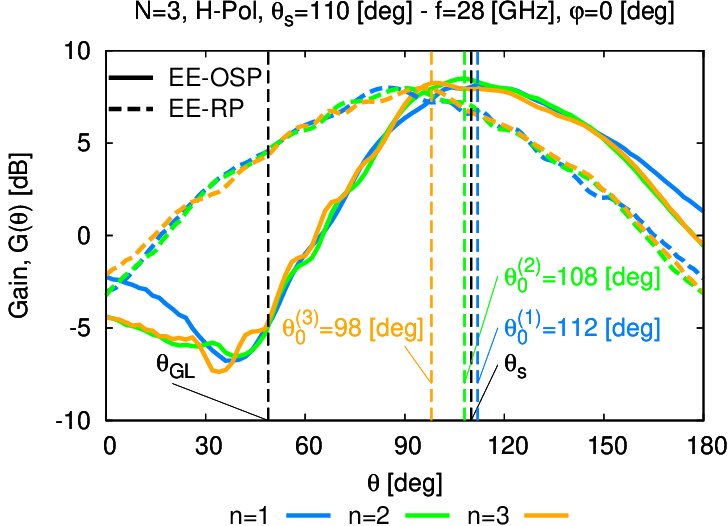}\end{center}

\begin{center}~\vfill\end{center}

\begin{center}\textbf{Fig. 8 - A. Benoni et} \textbf{\emph{al.}}\textbf{,}
\textbf{\emph{{}``}}Co-Design of Low-Profile Linear Microstrip Arrays
...''\end{center}

\newpage
\begin{center}~\vfill\end{center}

\begin{center}\includegraphics[%
  width=0.90\columnwidth]{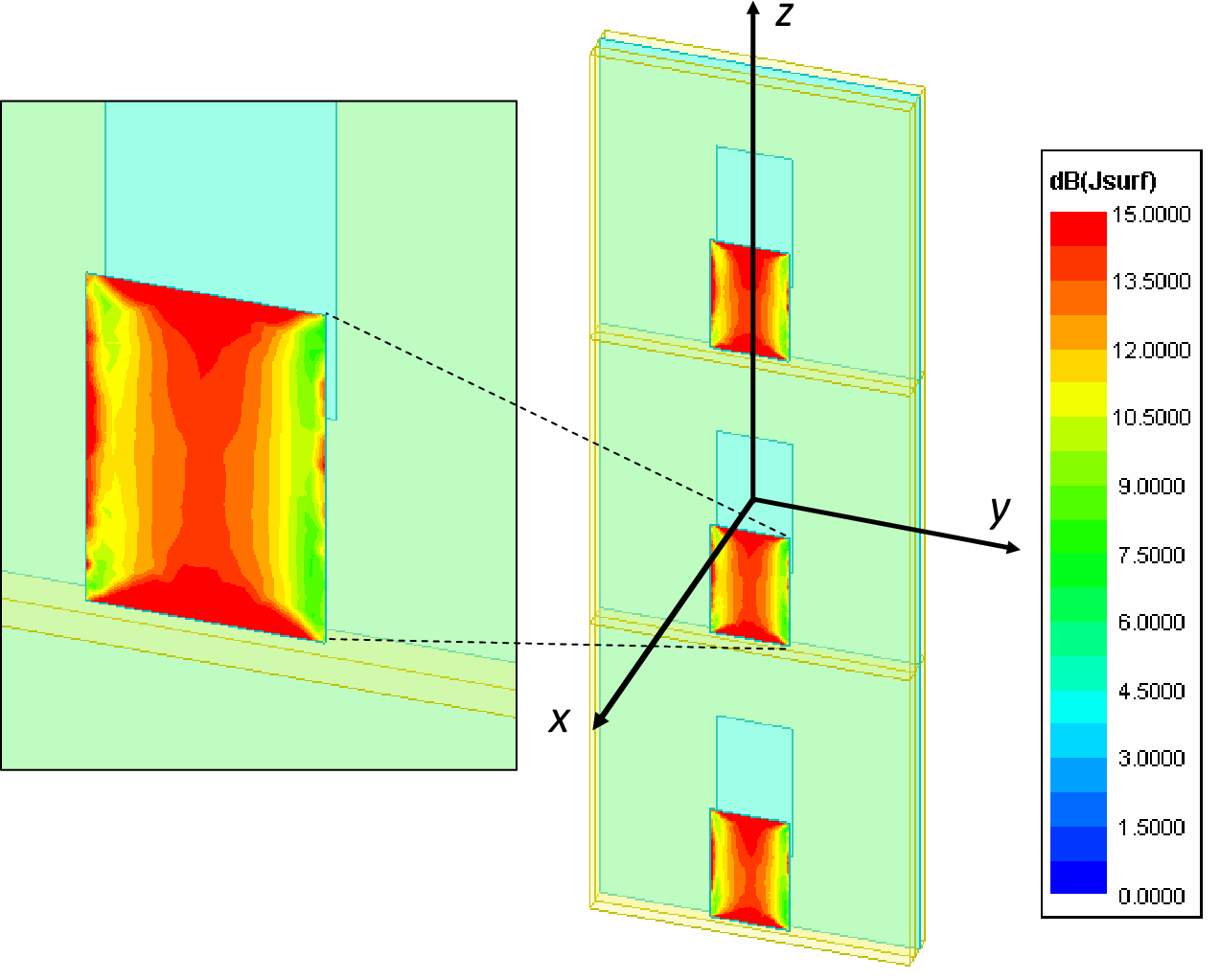}\end{center}

\begin{center}~\vfill\end{center}

\begin{center}\textbf{Fig. 9 - A. Benoni et} \textbf{\emph{al.}}\textbf{,}
\textbf{\emph{{}``}}Co-Design of Low-Profile Linear Microstrip Arrays
...''\end{center}

\newpage
\begin{center}~\vfill\end{center}

\begin{center}\includegraphics[%
  width=0.90\columnwidth]{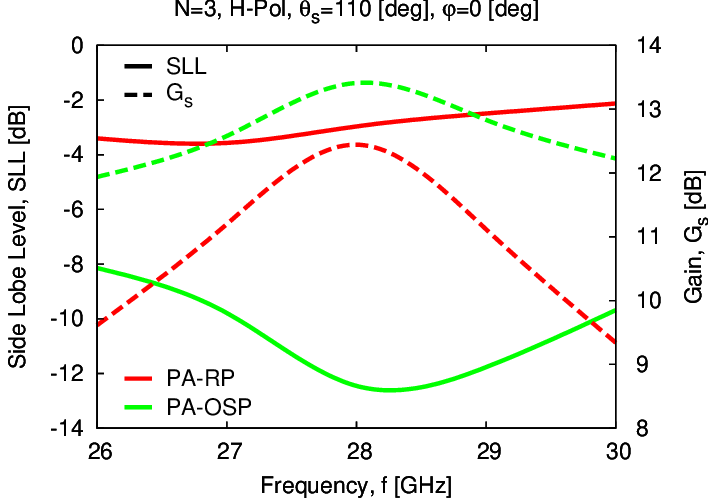}\end{center}

\begin{center}~\vfill\end{center}

\begin{center}\textbf{Fig. 10 - A. Benoni et} \textbf{\emph{al.}}\textbf{,}
\textbf{\emph{{}``}}Co-Design of Low-Profile Linear Microstrip Arrays
...''\end{center}

\newpage
\begin{center}~\vfill\end{center}

\begin{center}\begin{tabular}{cc}
\includegraphics[%
  width=0.50\columnwidth]{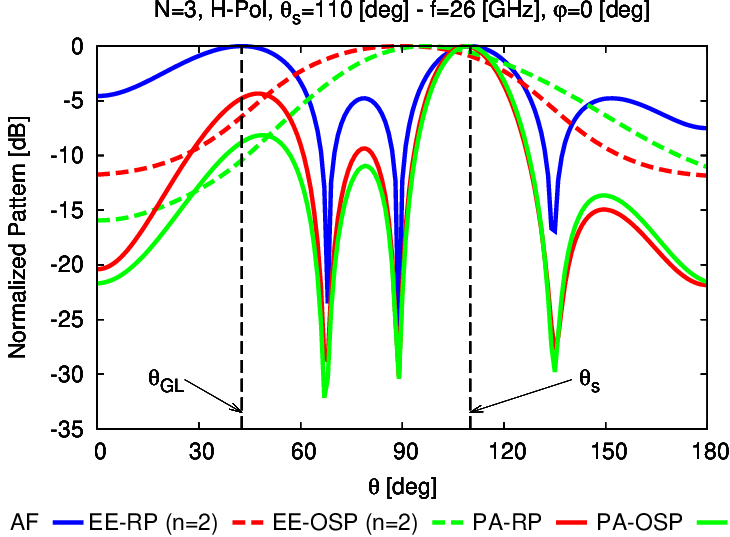}&
\includegraphics[%
  width=0.50\columnwidth]{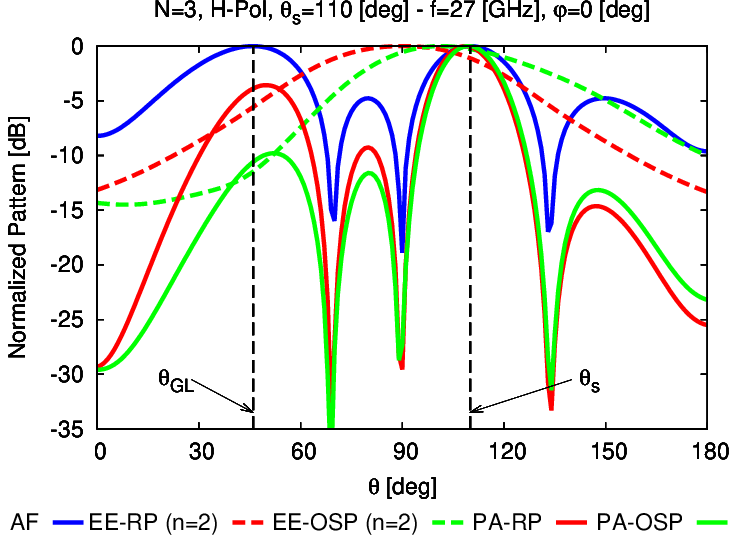}\tabularnewline
(\emph{a})&
(\emph{b})\tabularnewline
&
\tabularnewline
\includegraphics[%
  width=0.50\columnwidth]{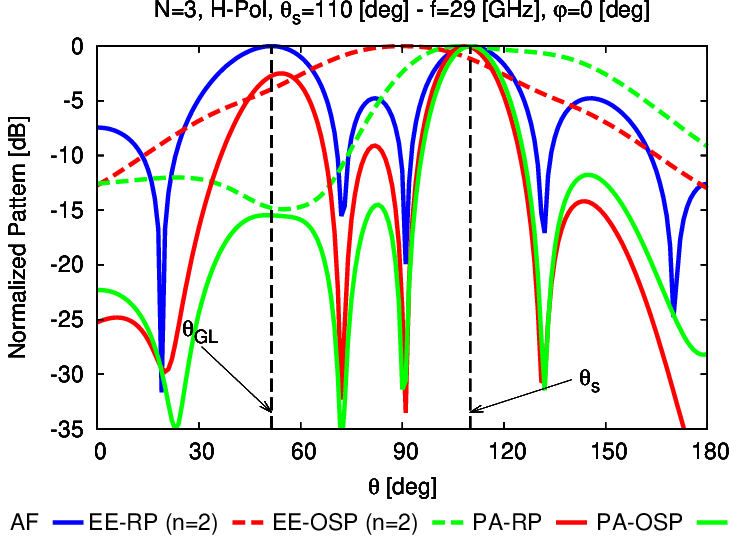}&
\includegraphics[%
  width=0.50\columnwidth]{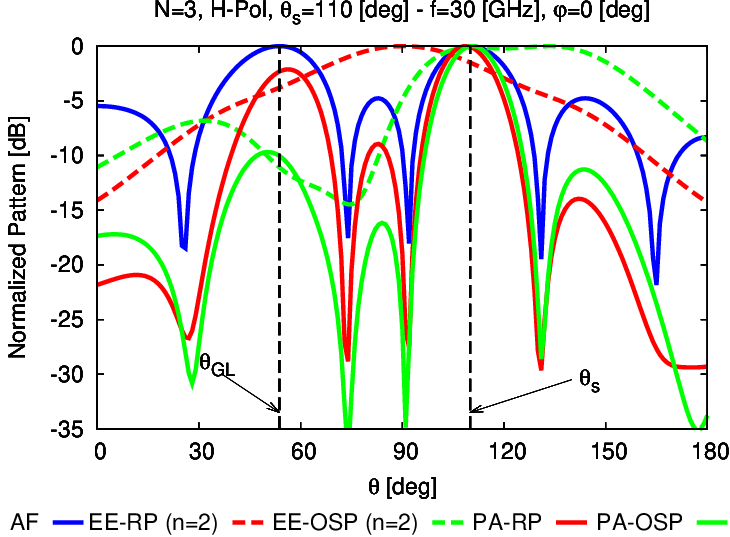}\tabularnewline
(\emph{c})&
(\emph{d})\tabularnewline
\end{tabular}\end{center}

\begin{center}~\vfill\end{center}

\begin{center}\textbf{Fig. 11 - A. Benoni et} \textbf{\emph{al.}}\textbf{,}
\textbf{\emph{{}``}}Co-Design of Low-Profile Linear Microstrip Arrays
...''\end{center}

\newpage
\begin{center}~\vfill\end{center}

\begin{center}\begin{tabular}{c}
\includegraphics[%
  width=0.75\columnwidth]{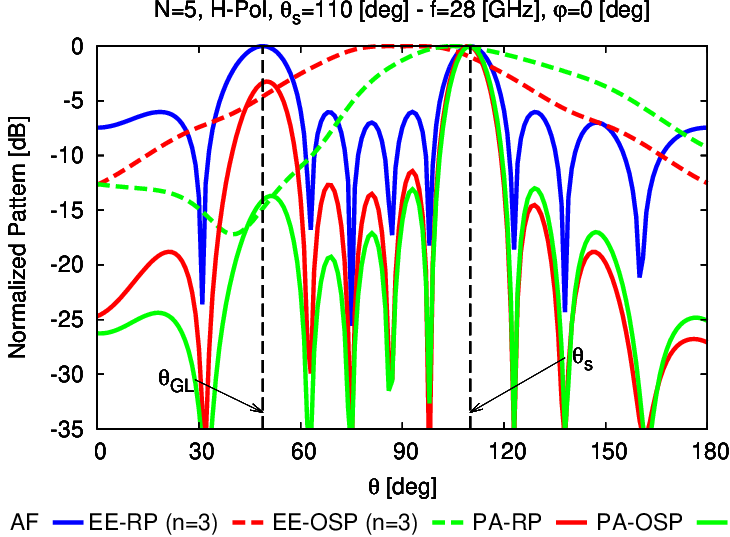}\tabularnewline
(\emph{a})\tabularnewline
\tabularnewline
\includegraphics[%
  width=0.75\columnwidth]{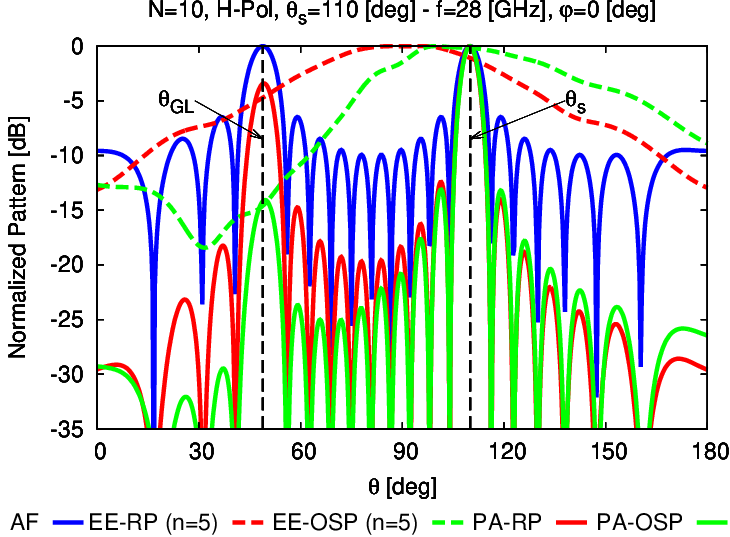}\tabularnewline
(\emph{b})\tabularnewline
\end{tabular}\end{center}

\begin{center}~\vfill\end{center}

\begin{center}\textbf{Fig. 12 - A. Benoni et} \textbf{\emph{al.}}\textbf{,}
\textbf{\emph{{}``}}Co-Design of Low-Profile Linear Microstrip Arrays
...''\end{center}

\newpage
\begin{center}~\vfill\end{center}

\begin{center}\includegraphics[%
  width=0.90\columnwidth]{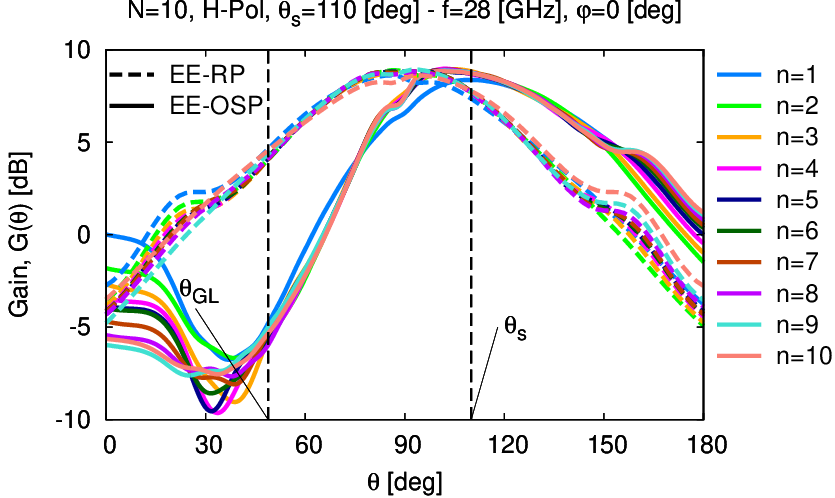}\end{center}

\begin{center}~\vfill\end{center}

\begin{center}\textbf{Fig. 13 - A. Benoni et} \textbf{\emph{al.}}\textbf{,}
\textbf{\emph{{}``}}Co-Design of Low-Profile Linear Microstrip Arrays
...''\end{center}

\newpage
\begin{center}~\vfill\end{center}

\begin{center}\begin{tabular}{cc}
\multicolumn{2}{c}{\includegraphics[%
  width=0.75\columnwidth]{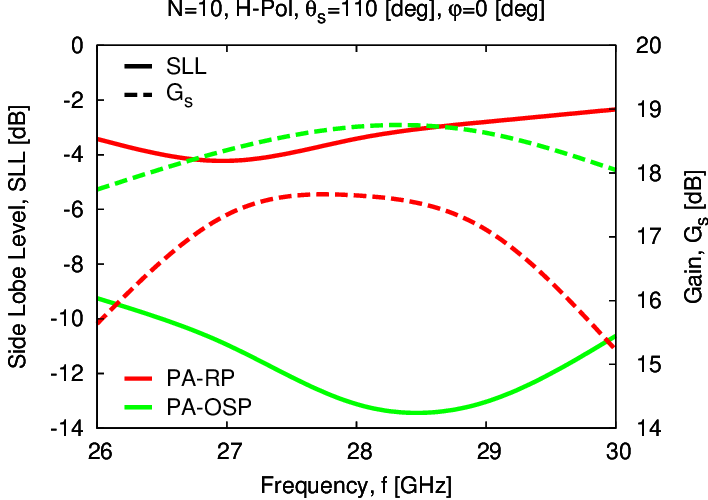}}\tabularnewline
\multicolumn{2}{c}{(\emph{a})}\tabularnewline
\multicolumn{2}{c}{}\tabularnewline
\multicolumn{2}{c}{\includegraphics[%
  width=0.75\columnwidth]{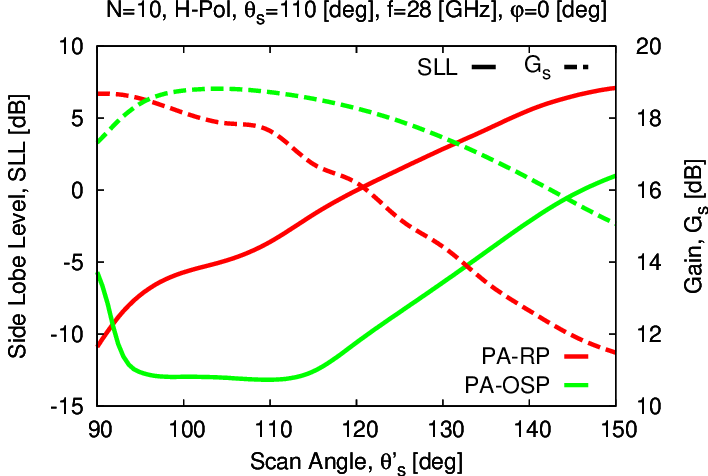}}\tabularnewline
\multicolumn{2}{c}{(\emph{b})}\tabularnewline
\end{tabular}\end{center}

\begin{center}~\vfill\end{center}

\begin{center}\textbf{Fig. 14 - A. Benoni et} \textbf{\emph{al.}}\textbf{,}
\textbf{\emph{{}``}}Co-Design of Low-Profile Linear Microstrip Arrays
...''\end{center}

\newpage
\begin{center}\begin{tabular}{c}
\includegraphics[%
  width=0.45\paperwidth]{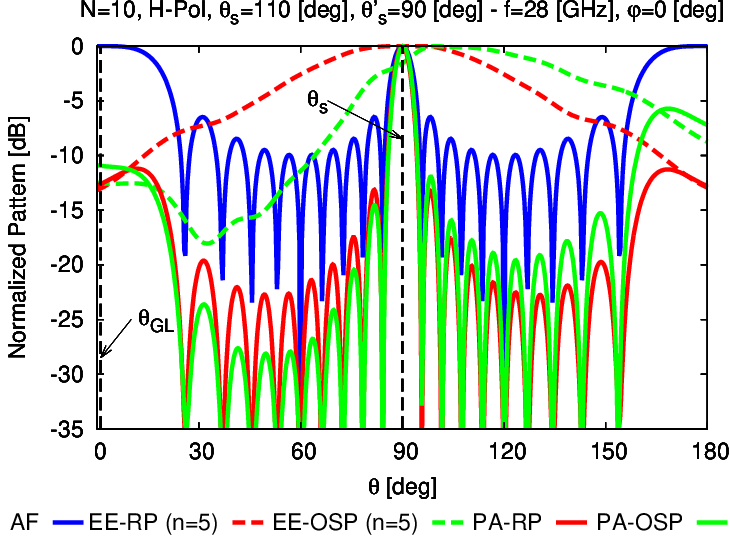}\tabularnewline
(\emph{a})\tabularnewline
\includegraphics[%
  width=0.45\paperwidth]{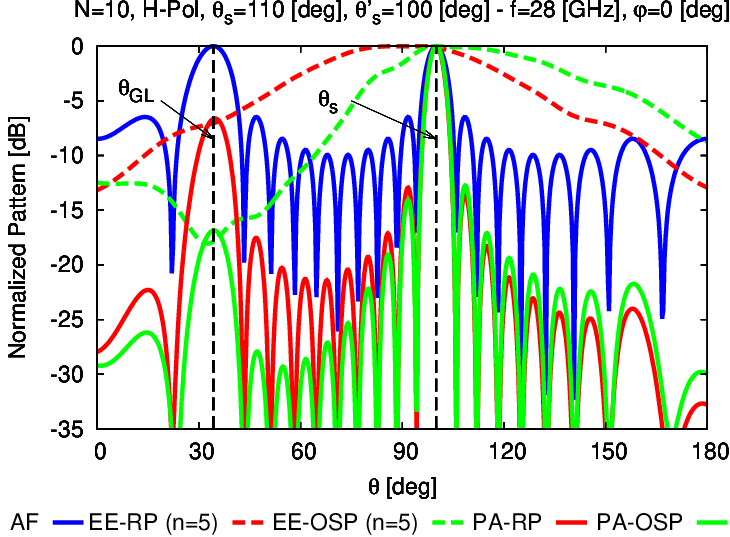}\tabularnewline
(\emph{b})\tabularnewline
\multicolumn{1}{c}{\includegraphics[%
  width=0.45\paperwidth]{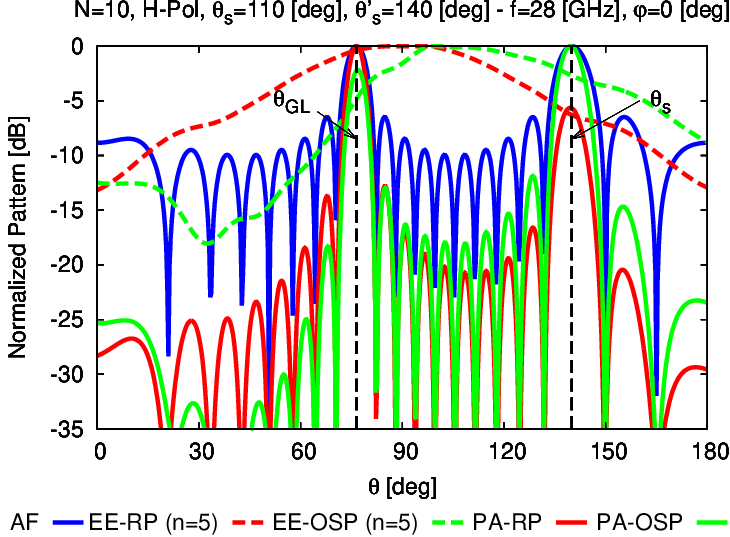}}\tabularnewline
\multicolumn{1}{c}{(\emph{c})}\tabularnewline
\end{tabular}\end{center}

\begin{center}\textbf{Fig. 15 - A. Benoni et} \textbf{\emph{al.}}\textbf{,}
\textbf{\emph{{}``}}Co-Design of Low-Profile Linear Microstrip Arrays
...''\end{center}

\newpage
\begin{center}~\vfill\end{center}

\begin{center}\includegraphics[%
  width=0.65\columnwidth]{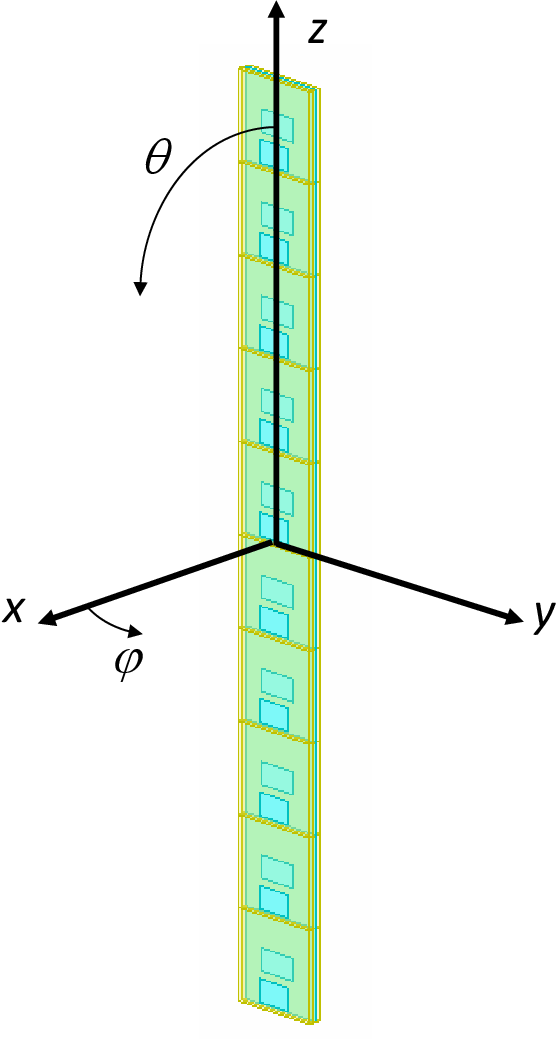}\end{center}

\begin{center}~\vfill\end{center}

\begin{center}\textbf{Fig. 16 - A. Benoni et} \textbf{\emph{al.}}\textbf{,}
\textbf{\emph{{}``}}Co-Design of Low-Profile Linear Microstrip Arrays
...''\end{center}

\newpage
\begin{center}~\vfill\end{center}

\begin{center}\begin{tabular}{c}
\includegraphics[%
  width=0.80\columnwidth]{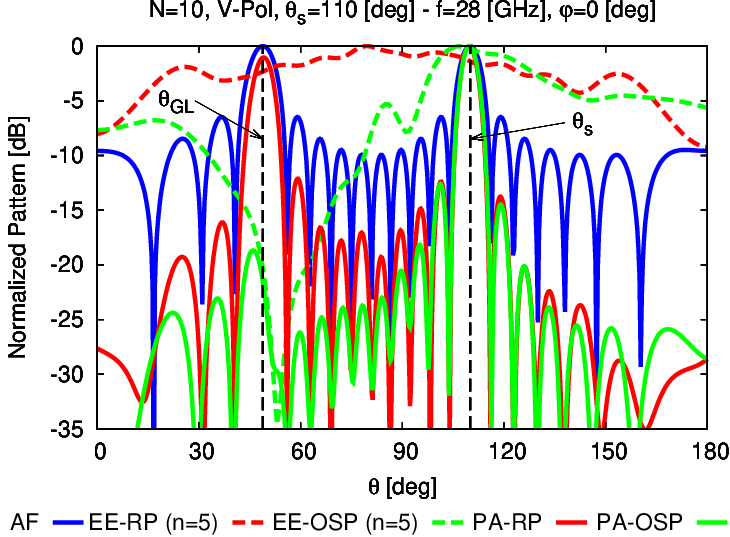}\tabularnewline
(\emph{a})\tabularnewline
\tabularnewline
\includegraphics[%
  width=0.80\columnwidth]{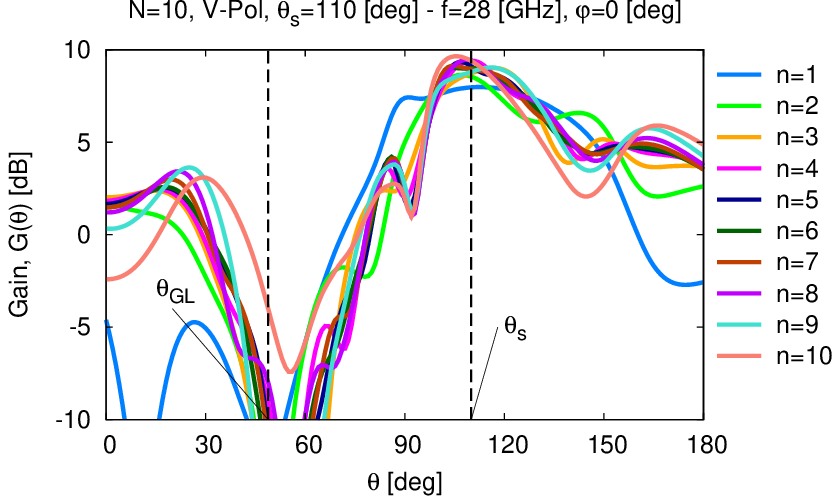}\tabularnewline
(\emph{b})\tabularnewline
\end{tabular}\end{center}

\begin{center}~\vfill\end{center}

\begin{center}\textbf{Fig. 17 - A. Benoni et} \textbf{\emph{al.}}\textbf{,}
\textbf{\emph{{}``}}Co-Design of Low-Profile Linear Microstrip Arrays
...''\end{center}

\newpage
\begin{center}~\vfill\end{center}

\begin{center}\includegraphics[%
  width=0.90\columnwidth]{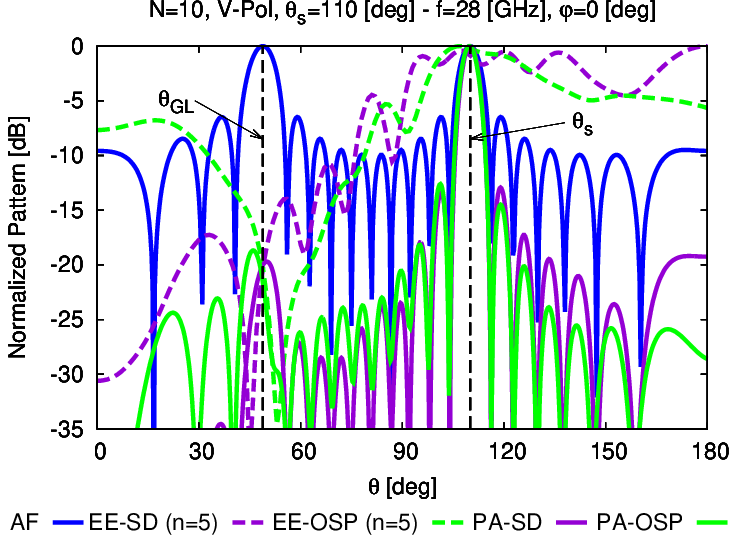}\end{center}

\begin{center}~\vfill\end{center}

\begin{center}\textbf{Fig. 18 - A. Benoni et} \textbf{\emph{al.}}\textbf{,}
\textbf{\emph{{}``}}Co-Design of Low-Profile Linear Microstrip Arrays
...''\end{center}

\newpage
\begin{center}~\vfill\end{center}

\begin{center}\includegraphics[%
  width=0.90\columnwidth]{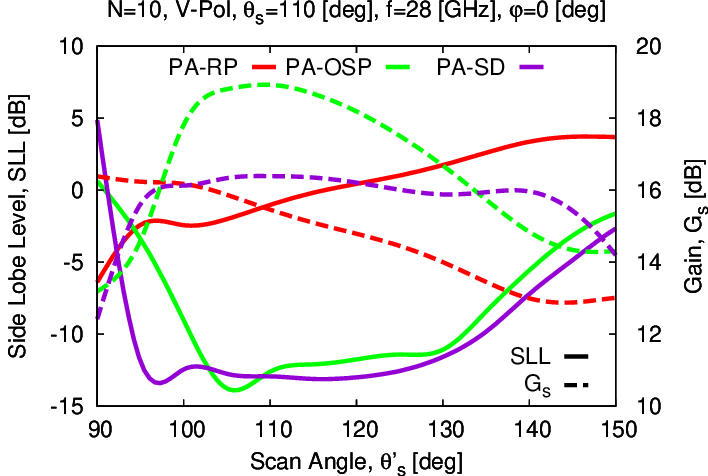}\end{center}

\begin{center}~\vfill\end{center}

\begin{center}\textbf{Fig. 19 - A. Benoni et} \textbf{\emph{al.}}\textbf{,}
\textbf{\emph{{}``}}Co-Design of Low-Profile Linear Microstrip Arrays
...''\end{center}

\newpage
\begin{center}~\vfill\end{center}

\begin{center}\begin{tabular}{cc}
\includegraphics[%
  width=0.45\columnwidth]{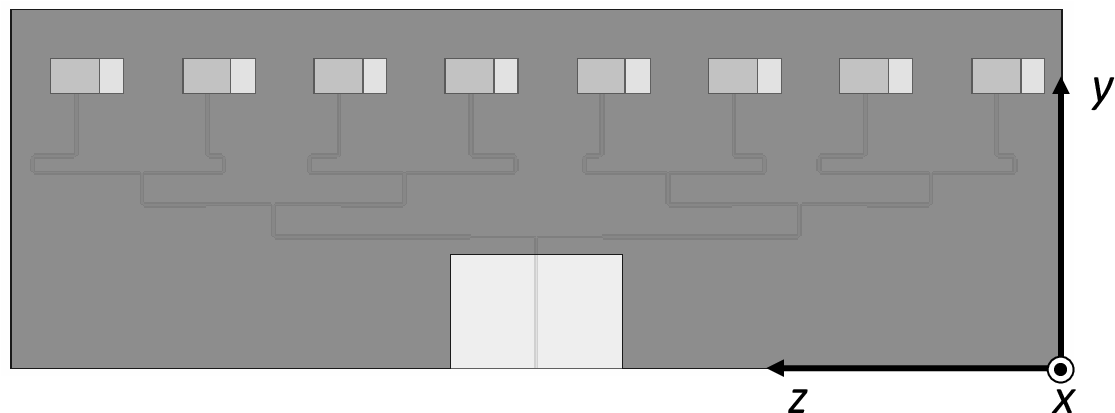}&
\includegraphics[%
  width=0.44\columnwidth]{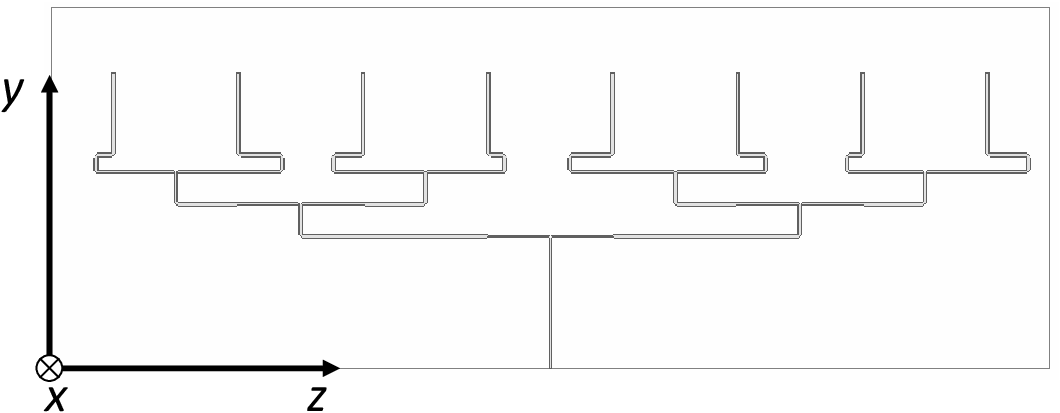}\tabularnewline
(\emph{a})&
(\emph{b})\tabularnewline
&
\tabularnewline
\includegraphics[%
  width=0.45\columnwidth]{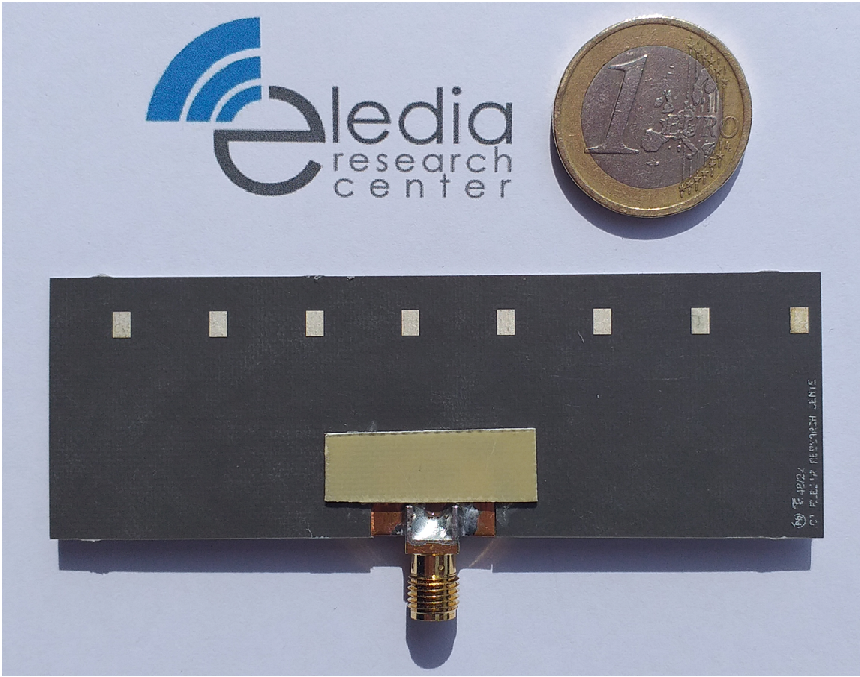}&
\includegraphics[%
  width=0.45\columnwidth]{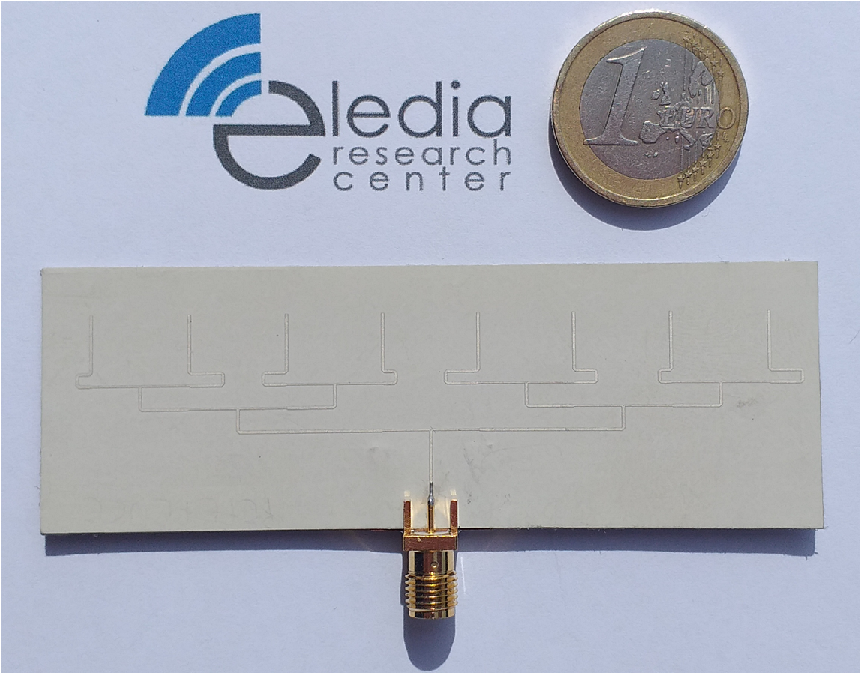}\tabularnewline
(\emph{c})&
(\emph{d})\tabularnewline
\end{tabular}\end{center}

\begin{center}~\vfill\end{center}

\begin{center}\textbf{Fig. 20 - A. Benoni et} \textbf{\emph{al.}}\textbf{,}
\textbf{\emph{{}``}}Co-Design of Low-Profile Linear Microstrip Arrays
...''\end{center}

\newpage
\noindent \begin{center}~\vfill\end{center}

\begin{center}\begin{tabular}{c}
\includegraphics[%
  width=0.75\columnwidth]{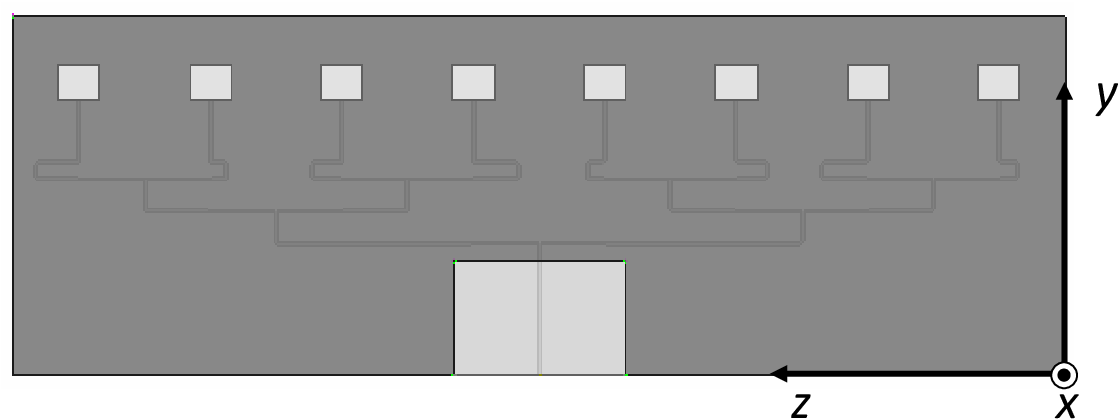}\tabularnewline
(\emph{a})\tabularnewline
\tabularnewline
\includegraphics[%
  width=0.75\columnwidth]{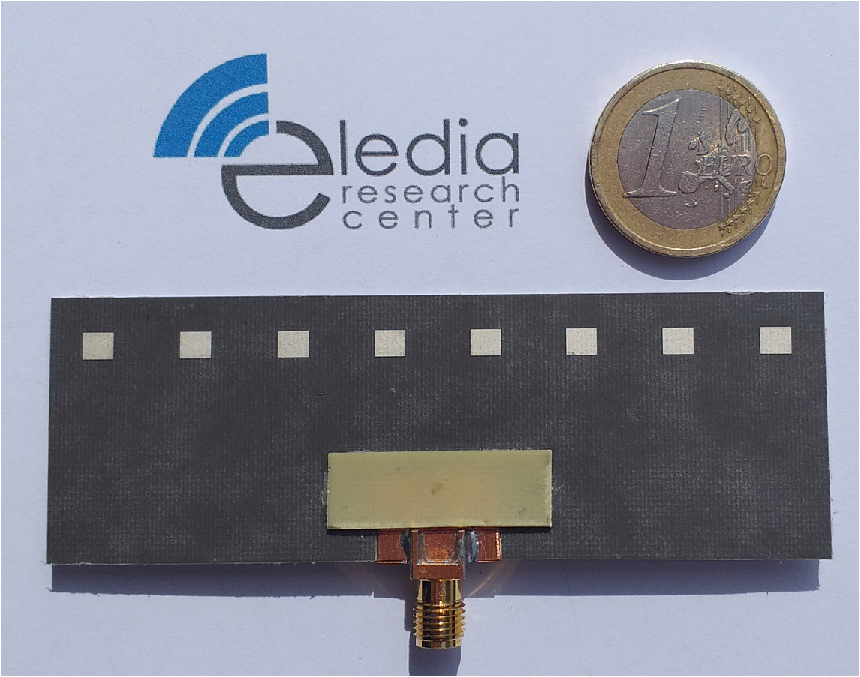}\tabularnewline
(\emph{b})\tabularnewline
\end{tabular}\end{center}

\begin{center}~\vfill\end{center}

\begin{center}\textbf{Fig. 21 - A. Benoni et} \textbf{\emph{al.}}\textbf{,}
\textbf{\emph{{}``}}Co-Design of Low-Profile Linear Microstrip Arrays
...''\end{center}

\newpage
\begin{center}~\vfill\end{center}

\begin{center}\includegraphics[%
  width=0.75\columnwidth]{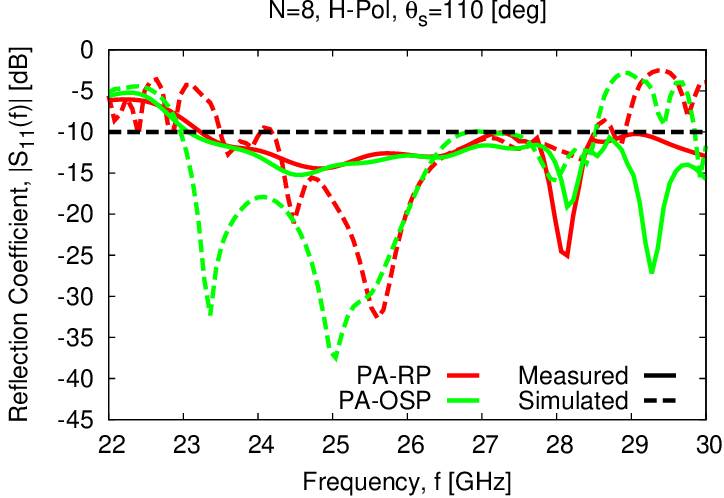}\end{center}

\begin{center}~\vfill\end{center}

\begin{center}\textbf{Fig. 22 - A. Benoni et} \textbf{\emph{al.}}\textbf{,}
\textbf{\emph{{}``}}Co-Design of Low-Profile Linear Microstrip Arrays
...''\end{center}

\newpage
\begin{center}~\vfill\end{center}

\begin{center}\begin{tabular}{c}
\includegraphics[%
  width=0.75\columnwidth]{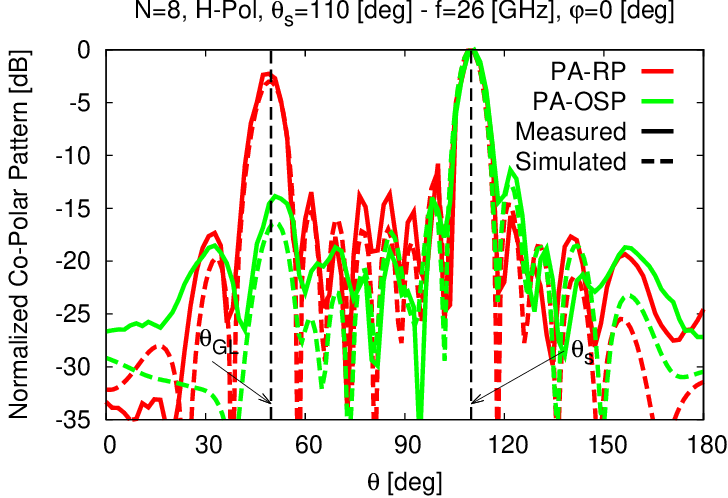}\tabularnewline
(\emph{a})\tabularnewline
\tabularnewline
\includegraphics[%
  width=0.75\columnwidth]{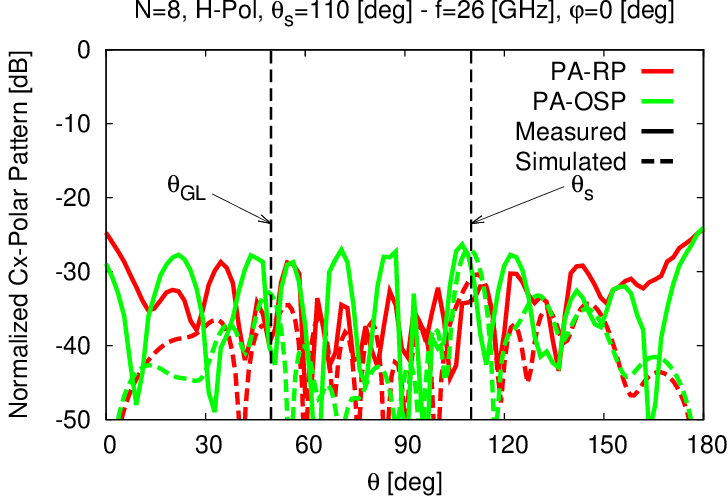}\tabularnewline
(\emph{b})\tabularnewline
\end{tabular}\end{center}

\begin{center}~\vfill\end{center}

\begin{center}\textbf{Fig. 23 - A. Benoni et} \textbf{\emph{al.}}\textbf{,}
\textbf{\emph{{}``}}Co-Design of Low-Profile Linear Microstrip Arrays
...''\end{center}

\newpage
\begin{center}~\vfill\end{center}

\begin{center}\begin{tabular}{|c|c||c||c|}
\hline 
$k$&
\emph{DoF}&
\emph{H-Pol} {[}m{]}&
\emph{V-Pol} {[}m{]}\tabularnewline
&
&
(Fig. 5)&
(Fig. 16) \tabularnewline
\hline
\hline 
$1$&
$L_{f}$&
$4.63\times10^{-3}$&
$4.63\times10^{-3}$\tabularnewline
\hline
$2$&
$W_{f}$&
$3.14\times10^{-4}$&
$3.56\times10^{-4}$\tabularnewline
\hline
$3$&
$L_{s}$&
$2.42\times10^{-3}$&
$2.43\times10^{-3}$\tabularnewline
\hline
$4$&
$W_{s}$&
$4.80\times10^{-4}$&
$4.79\times10^{-4}$\tabularnewline
\hline
$5$&
$O_{s}$&
$7.27\times10^{-4}$&
$7.26\times10^{-4}$\tabularnewline
\hline
$6$&
$L_{p}$&
$2.54\times10^{-3}$&
$2.64\times10^{-3}$\tabularnewline
\hline
$7$&
$W_{p}$&
$4.83\times10^{-3}$&
$4.86\times10^{-3}$\tabularnewline
\hline
$8$&
$L_{d}$&
$2.63\times10^{-3}$&
$2.70\times10^{-3}$\tabularnewline
\hline
$9$&
$W_{d}$&
$4.05\times10^{-3}$&
$4.46\times10^{-3}$\tabularnewline
\hline
$10$&
$\Delta_{z}$&
$2.83\times10^{-3}$&
$3.35\times10^{-3}$\tabularnewline
\hline
\end{tabular}\end{center}

\begin{center}~\vfill\end{center}

\begin{center}\textbf{Tab. I - A. Benoni et} \textbf{\emph{al.}}\textbf{,}
\textbf{\emph{{}``}}Co-Design of Low-Profile Microstrip Arrays ...''\end{center}

\newpage
\begin{center}~\vfill\end{center}

\begin{center}\begin{tabular}{|c||c||c||c||c|c|}
\hline 
\emph{Test Case}&
\emph{Polarization}&
$N$&
\emph{Antenna}&
$SLL$ {[}dB{]}&
$G_{s}\ $ {[}dB{]}\tabularnewline
\hline
\hline 
Fig. 6(\emph{a})&
\emph{H}-Pol&
$3$&
\emph{PA-RP }&
$-2.96\ $&
$12.44$\tabularnewline
\hline
Fig. 6(\emph{a})&
\emph{H}-Pol&
$3$&
\emph{PA-OSP}&
$-12.45\ $&
$13.41$\tabularnewline
\hline
\hline 
Fig. 12(\emph{a})&
\emph{H}-Pol&
$5$&
\emph{PA-RP }&
$-3.23$&
$14.60$\tabularnewline
\hline
Fig. 12(\emph{a})&
\emph{H}-Pol&
$5$&
\emph{PA-OSP}&
$-13.00$&
$15.60$\tabularnewline
\hline
\hline 
Fig. 12(\emph{b})&
\emph{H}-Pol&
$10$&
\emph{PA-RP }&
$-3.60$&
$17.65$\tabularnewline
\hline
Fig. 12(\emph{b})&
\emph{H}-Pol&
$10$&
\emph{PA-OSP}&
$-14.57$&
$18.72$\tabularnewline
\hline
\hline 
Fig. 17(\emph{a})&
\emph{V}-Pol&
$10$&
\emph{PA-RP}&
$-1.02$&
$15.43$\tabularnewline
\hline
Fig. 17(\emph{a})&
\emph{V}-Pol&
$10$&
\emph{PA-OSP}&
$-12.85$&
$18.92$\tabularnewline
\hline
Fig. 18&
\emph{V-}Pol&
$10$&
\emph{PA-SD} \cite{Yepes 2020.b}&
$-12.93$&
$16.39$\tabularnewline
\hline
\end{tabular}\end{center}

\begin{center}~\vfill\end{center}

\begin{center}\textbf{Tab. II - A. Benoni et} \textbf{\emph{al.}}\textbf{,}
\textbf{\emph{{}``}}Co-Design of Low-Profile Microstrip Arrays ...''\end{center}

\begin{thebibliography}{10}
\bibitem{Yepes 2020.b}C. Yepes, E. Gandini, S. Monni, A. Neto, F. E. van Vliet, and D. Cavallo,
{}``A linear array of skewed dipoles with asymmetric radiation pattern
for angular filtering,'' \emph{IEEE Antennas Wireless Propag. Lett}.,
vol. 19, no. 3, pp. 408-412, Mar. 2020.
\bibitem{Kim 2015}I. Kim and Y. Rahmat-Samii, {}``Electromagnetic band gap-dipole sub-array
antennas creating an enhanced tilted beams for future base station,''
\emph{IET Microw. Antennas Propag.}, vol. 9, no. 4, pp. 319-327, Mar.
2015.
\bibitem{Ramaccia 2022}D. Ramaccia, M. Barbuto, A. Monti, S. Vellucci, C. Massagrande, A.
Toscano, and F. Bilotti, {}``Metasurface dome for above-the-horizon
grating lobes reduction in 5G-NR systems,'' \emph{IEEE Antennas Wireless
Propag. Lett.}, 2022, vol. 21, no. 11, pp. 2176-2180, Nov. 2022.
\bibitem{Rocca 2016}P. Rocca, G. Oliveri, R. J. Mailloux, and A. Massa, {}``Unconventional
phased array architectures and design methodologies - A review,''
\emph{Proc. IEEE}, vol. 104, no. 3, pp. 544-560, Mar. 2016.
\bibitem{Balanis 2016}C. Balanis, \emph{Antenna theory: analysis and design}. Hoboken, N.J.:
Wiley, 2016.
\bibitem{Herd 2016}J. S. Herd and M. David Conway, {}``The evolution to modern phased
array architectures,'' \emph{Proc. IEEE}, vol. 104, no. 3, pp. 519-529,
2016.
\bibitem{Mailloux 2018}R. J. Mailloux, \emph{Phased Array Antenna Handbook} (3rd ed). Norwood,
MA, USA: Artech House, 2018.
\bibitem{Cox 2021}C. Cox, \emph{An Introduction to 5G: The New Radio, 5G Network and
Beyond}. Hoboken, NJ, USA: John Wiley \& Sons, 2021.
\bibitem{Hataria 2021}H. Tataria, M. Shafi, A. F. Molisch, M. Dohler, H. Sjoland, and F.
Tufvesson, {}``6G wireless systems: Vision, requirements, challenges,
insights, and opportunities,'' \emph{Proc. IEEE}, vol. 109, no. 7,
pp. 1166-1199, Jul. 2021.
\bibitem{Blanco 2014}D. Blanco, N. Llombart, and E. Rajo-Iglesias, {}``On the use of leaky
wave phased arrays for the reduction of the grating lobe level,''
\emph{IEEE Trans. Antennas Propag.}, vol. 62, no. 4, pp. 1789-1795,
Apr. 2014.
\bibitem{Blanco 2016}D. Blanco, E. Rajo-Iglesias, A. M. Benito, and N. Llombart, {}``Leaky-wave
thinned phased array in PCB technology for telecommunication applications,''
\emph{IEEE Trans. Antennas Propag.}, vol. 64, no. 10, pp. 4288-4296,
Oct. 2016.
\bibitem{Jiang 2017}M. Jiang, Z. N. Chen, Y. Zhang, W. Hong, and X. Xuan, {}``Metamaterial-based
thin planar lens antenna for spatial beamforming and multibeam massive
MIMO,'' \emph{IEEE Trans. Antennas Propag}., vol. 65, no. 2, pp.
464-472, Feb. 2017.
\bibitem{Ye 2019}L. H. Ye, X. Y. Zhang, Y. Gao, and Q. Xue, {}``Wideband dual-polarized
two-beam antenna array with low sidelobe and grating-lobe levels for
base-station applications,'' \emph{IEEE Trans. Antennas Propag.},
vol. 67, no. 8, pp. 5334-5343, Aug. 2019.
\bibitem{Yepes 2020.a}C. Yepes, E. Gandini, S. Monni, A. Neto, F. E. van Vliet, and D. Cavallo,
{}``Analysis of tilted dipole arrays: Impedance and radiation properties,''
\emph{IEEE Trans. Antennas Propag.}, vol. 68, no. 1, pp. 254-265,
Jan. 2020.
\bibitem{Lee 2003}Y. J. Lee, S. H. Jeong, W. S. Park, J. S. Yun, and S. I. Jeon, {}``Multilayer
spatial angular filter with airgap tuners to suppress grating lobes
of $4\times1$ array antenna,'' \emph{Electron. Lett.}, vol. 39,
no. 1, pp. 15-17, Jan. 2003.
\bibitem{Lee 2005}Y. J. Lee, J. Yeo, R. Mittra, and W. S. Park, {}``Application of
electromagnetic bandgap (EBG) superstrates with controllable defects
for a class of patch antennas as spatial angular filters,'' \emph{IEEE
Trans. Antennas Propag}., vol. 53, no. 1, pp. 224-235, Jan. 2005.
\bibitem{Iqbal 2018}Z. Iqbal and M. Pour, {}``Grating lobe reduction in scanning phased
array antennas with large element spacing,'' \emph{IEEE Trans. Antennas
Propag.}, vol. 66, no. 12, pp. 6965-6974, Dec. 2018.
\bibitem{Iqbal 2019}Z. Iqbal and M. Pour, {}``Exploiting higher order modes for grating
lobe reduction in scanning phased array antennas,'' \emph{IEEE Trans.
Antennas Propag.}, vol. 67, no. 11, pp. 7144-7149, Nov. 2019.
\bibitem{Rajo-Iglesias 2002}E. Rajo-Iglesias, G. Villaseca-Sanchez, and C. Martin-Pascual, {}``Input
impedance behavior in offset stacked patches,'' \emph{IEEE Antennas
Wireless Propag. Lett.}, vol. 1, pp. 28-30, 2002.
\bibitem{Sarin 2011}V. P. Sarin, M. S. Nishamol, D. Tony, C. K. Aanandan, P. Mohanan,
and K. Vasudevan, {}``A wideband stacked offset microstrip antenna
with Improved gain and low cross polarization,'' \emph{IEEE Trans.
Antennas Propag.}, vol. 59, no. 4, pp. 1376-1379, Apr. 2011.
\bibitem{Katyal 2017.a}A. Katyal and A. Basu, {}``Analysis and optimisation of broadband
stacked microstrip antennas using transmission line model,'' \emph{IET
Microw. Antennas Propag.}, vol. 11, no. 1, pp. 81-91, Jan. 2017.
\bibitem{Katyal 2017.b}A. Katyal and A. Basu, {}``Compact and broadband stacked microstrip
patch antenna for target scanning applications,'' \emph{IEEE Antennas
Wireless Propag. Lett.}, vol. 16, pp. 381-384, 2017.
\bibitem{Luk 1993}K. M. Luk, K. F. Tong, and T. M. Au, {}``Offset dual-patch microstrip
antenna,'' \emph{Electron. Lett.}, vol. 29, no. 18, pp. 1635-1636,
Sep. 1993.
\bibitem{Rajo-Iglesias 2003}E. Rajo-Iglesias, J.-L. Vazquez-Roy, L. Inclan-Sanchez, D. Segovia-Vargas,
V. Gonzalez-Posadas, and C. Martin-Pascual, {}``Offset stacked patches
behavior in an array,'' \emph{Microw. Opt. Technol. Lett.}, vol.
40, no. 3, pp. 262-265, Dec. 2003.
\bibitem{Oliveri 2017b}G. Oliveri, G. Gottardi, F. Robol, A. Polo, L. Poli, M. Salucci, M.
Chuan, C. Massagrande, P. Vinetti, M. Mattivi, R. Lombardi, and A.
Massa, {}``Codesign of unconventional array architectures and antenna
elements for 5G base stations,'' \emph{IEEE Trans. Antennas Propag.},
vol. 65, no. 12, pp. 6752-6767, Dec. 2017.
\bibitem{Oliveri 2017}G. Oliveri, M. Salucci, N. Anselmi, and A. Massa, {}``Multiscale
system-by-design synthesis of printed WAIMs for waveguide array enhancement,''
\emph{IEEE J. Multiscale Multiphysics Computat. Techn.}, vol. 2, pp.
84-96, 2017.
\bibitem{Oliveri 2021}G. Oliveri, A. Polo, M. Salucci, G. Gottardi, and A. Massa, {}``SbD-Based
synthesis of low-profile WAIM superstrates for printed patch arrays,''
\emph{IEEE Trans. Antennas Propag.}, vol. 69, no. 7, pp. 3849-3862,
Jul. 2021.
\bibitem{Massa 2021b}A. Massa and M. Salucci, {}``On the design of complex \emph{EM} devices
and systems through the System-by-Design paradigm - A framework for
dealing with the computational complexity,'' \emph{IEEE Trans. Antennas
Propag.}, vol. 70, no. 2, pp. 1328-1343, Feb. 2022.
\bibitem{Salucci 2021}M. Salucci, G. Oliveri, M. A. Hannan, and A. Massa, {}``System-by-Design
paradigm-based synthesis of complex systems: The case of spline-contoured
3D radomes,'' \emph{IEEE Antennas Propag. Mag.}, vol. 64, no. 1,
pp. 72-83, Feb. 2022.
\bibitem{Pozar 1994}D. M. Pozar, {}``The active element pattern,'' \emph{IEEE Trans.
Antennas Propag.}, vol. 42, no. 8, pp. 1176-1178, Aug. 1994.
\bibitem{Rocca 2009}P. Rocca, M. Benedetti, M. Donelli, D. Franceschini, and A. Massa,
{}``Evolutionary optimization as applied to inverse scattering problems,''
\emph{Inv. Prob.}, vol. 24, pp. 1-41, 2009.
\bibitem{Goudos 2021}S. K. Goudos, \emph{Emerging Evolutionary Algorithms for Antennas
and Wireless Communications}. IET, 2021 (ISBN 1785615521).
\bibitem{Massa 2018b}A. Massa, G. Oliveri, M. Salucci, N. Anselmi, and P. Rocca, ''Learning-by-examples
techniques as applied to electromagnetics,'' \emph{J. Electromagn.
Waves Appl.}, vol. 32, no. 4, pp. 516-541, 2018.
\bibitem{Forrester 2008}A. I. J. Forrester, A. Sobester, and A. J. Keane, \emph{Engineering
Design via Surrogate Modelling: A Practical Guide}. Hoboken, N.J.:
John Wiley \& Sons, 2008.
\bibitem{Jones 1998}D. R. Jones, M. Schonlau, and W. J. Welch, {}``Efficient global optimization
of expensive black-box functions,'' \emph{J. Global Opt}., vol. 13,
pp. 455-492, 1998.
\bibitem{HFSS 2021}ANSYS Electromagnetics Suite - HFSS (2021). ANSYS, Inc.
\bibitem{Perlmutter 1985}P. Perlmutter, S. Shtrikman, and D. Treves, {}``Electric surface
current model for the analysis of microstrip antennas with application
to rectangular elements,'' \emph{IEEE Trans. Antennas Propag.}, vol.
33, no. 3, pp. 301-311, Mar. 1985.
\end{thebibliography}
\end{document}